\documentclass[letterpaper,journal]{IEEEtran}
\usepackage{amsmath,amsfonts}
\usepackage{algorithm}
\usepackage{array}
\usepackage[caption=false,font=normalsize,labelfont=sf,textfont=sf]{subfig}

\usepackage{textcomp}
\usepackage{stfloats}
\usepackage{url}
\usepackage{verbatim}
\usepackage{graphicx}
\usepackage{cite}
\usepackage{amsmath, amssymb}
\usepackage{graphicx}
\usepackage{algorithm,float}
\usepackage{algpseudocode}
\usepackage{bm}
\usepackage{tikz}

\newcommand{\revised}[1]{#1}

\begin{document}

\title{Lorentzian-Constrained Holographic Beamforming Optimization in Multi-user Networks with Dynamic Metasurface Antennas}

\author{Askin Altinoklu, and Leila Musavian
\thanks{The authors are with the School of Computer Science and Electronic Engineering, University of Essex, Wivenhoe Park, Colchester CO43SQ, United Kingdom (e-mail: {askin.altinoklu, leila.musavian}@essex.ac.uk). This work was supported by UK Research and Innovation under the UK government’s Horizon Europe funding guarantee through MSCA-DN SCION Project Grant Agreement No.101072375 [Grant Number: EP/X027201/1]. Part of this work was presented at the 2025 European Conference on Networks and Communications (EuCNC) \cite{aaltinoklu}.}}


\maketitle

\begin{abstract}
Dynamic metasurface antennas (DMAs) are promising alternatives to fully digital (FD) architectures, enabling hybrid beamforming via low-cost reconfigurable metasurfaces. In DMAs, holographic beamforming is achieved through tunable elements by Lorentzian-constrained holography (LCH), significantly reducing the need for radio-frequency (RF) chains and analog circuitry. However, the Lorentzian constraints and limited RF chains introduce a trade-off between reduced system complexity and beamforming performance, especially in dense network scenarios. This paper addresses resource allocation in multi-user multiple-input-single-output (MISO) networks under the Signal-to-Interference-plus-Noise Ratio (SINR) constraints, aiming to minimize total transmit power. We propose a holographic beamforming algorithm based on the Generalized Method of Lorentzian-Constrained Holography (GMLCH), which optimizes DMA weights, yielding flexibility for using various LCH techniques to tackle the aforementioned trade-offs. Building upon GMLCH, we further propose a new algorithm i.e., Adaptive Radius Lorentzian Constrained Holography (ARLCH), which achieves optimization of DMA weights with additional degree of freedom in a greater optimization space, and provides lower transmitted power, while improving scalability for higher number of users. Numerical results show that ARLCH reduces power consumption by over 20\% compared to benchmarks, with increasing effectiveness as the number of users grows.

\end{abstract}


\section{Introduction}
\label{sec:Intro}
\IEEEPARstart{M}{assive} multiple-input multiple-output (MIMO) technology has been one of the key actors of 5G and beyond wireless communication networks \cite{5G}. Recently, with the reveal of anticipated requirements for sixth-generation (6G) communication systems, massive MIMO is evolving into extremely large massive MIMO systems to address the substantial increases associated with key performance indicators of 6G \cite{6g_kpi1,6g_kpi2}. While massive MIMO has demonstrated remarkable robustness in real-world applications \cite{5g_MIMO}, the complexity and challenges associated with evolving conventional MIMO structures into large-scale antenna arrays have driven significant interest in emerging technologies, such as reconfigurable intelligent surfaces (RIS) and Holographic MIMO (HMIMO) with their prominent features of holographic beamforming and dynamic reconfigurability of individual array elements. These technologies aim to address the practical limitations of current systems with their advantages, including low manufacturing costs, low complexity of hardware architecture, and reduced power consumption \cite{hMIMO1,hMIMO2,hMIMO3}. One particular example of different types of architectures for HMIMO is called dynamic metasurface antennas (DMAs). DMAs consists of an array of metasurface elements which are excited by the reference waves supplied by a waveguide or microstrip \cite{em_dma1} and the objective waves are generated through the aperture via beamforming with the help of reconfigurability of the physical properties of metamaterial elements, such as tunable polarizability\cite{em_dma2}. This structure is already shown to be an effective solution for MIMO, when they are combined with digital precoders, where array of metasurfaces are grouped within the array of waveguides (microstrips) and each of them is connected to the digital precoders via single radio-frequency (RF) chain. This allows beamforming to be performed partially in digital and analog domain through tunable metasurface antennas, and yields low size, weight, and power-consuming structures for MIMO applications\cite {DMA1, exp_dma1}. 
\IEEEpubidadjcol

The fundamental distinction of DMA-based beamforming compared to conventional architectures, such as fully-digital (FD) and hybrid analog/digital beamforming, lies in the reconfigurability of each individual metamaterial element that is achieved by shifting the resonance frequency to tune the amplitude and phase required for beamforming\revised{\cite{exp_dma1,exp_dma2,review_ref1}}. Additional adjustments to the beamforming can be made by modifying the amplitude and phase shifts of reference waves guided by microstrips or waveguides, which are controlled by the digital precoders through the RF chains. In the DMA architecture, the digital precoders are connected to the group of metasurface elements, allowing the number of required RF chains and power amplifiers to be reduced to match the number of microstrips or waveguides rather than the individual antenna elements. This feature of DMA based architectures offers low complexity for the transmitter side, especially for large-scale arrays. On the other hand, the reduction in the number of RF chains can dictate the decrease in the degree of freedom (DoF) in terms of independent beamforming parameters \cite{aaltinoklu}. The performance of FD architectures can be considered as an optimal benchmark for evaluating DMA-based architectures,  with the performance gap between FD and DMA architectures presenting the trade-off between reducing the number of RF chains and achieving optimal beamforming performance~\cite{DMA1, aaltinoklu, metasurface_6g}. Despite concerns about DoF, DMA-based beamforming architectures have so far demonstrated strong performance compared to FD architectures in various applications. These include downlink and uplink wireless information transfer (WIT) systems, optimizing parameters such as achievable sum-rate, spectral efficiency, and channel capacity \cite{dma_adv1, DMA_app1, DMA_BF1, DMA_app5}, as well as RF wireless power transfer (WPT) systems, optimizing RF harvested energy and transmitter power consumption \cite{DMA_app3, DMA_app4, Azarbahram_2024}. 
\revised{In addition, recent studies have investigated reconfigurable holographic surfaces under more practical conditions, showing robustness in terms of energy and spectral efficiency even when realistic hardware impairments such as transceiver non-idealities, power amplifier nonlinearities, and RF chain distortions are taken into account \cite{hwi_1,hwi_2}.}

In works \cite{dma_adv1, DMA_app1, DMA_BF1, DMA_app5,DMA_app3, DMA_app4, Azarbahram_2024}, different \revised{approaches} have been proposed for obtaining beamforming in DMA-assisted digital precoding setups for single user and multi-user communication networks.  These approaches include the alternating optimization of digital precoding vectors, and analog DMA weights, which are representing tunability of metasurfaces. A unique characteristic of metasurfaces, known as Lorentzian resonance, enabling analog-domain beamforming with DMA, is also a challenging aspect for the implementation of resource allocation algorithms in wireless communication networks, yet to be explored in terms of the drawbacks and benefits associated with the tunability of metasurfaces. In conventional architectures, beamforming is achieved by aligning the phases of array elements relative to the target direction, while maintaining constant amplitudes across the aperture. Consequently, the ideal weights for beamforming adhere to the structure of a constant-modulus complex circle.
\revised{On the other hand, in DMA systems, the metasurface elements behave as Lorentzian resonators, where the amplitude and phase responses are intrinsically coupled through a single tunable parameter determined by the element’s resonant electrical length and resonance frequency~\cite{em_dma2}. Physically, this resonant electrical length fixes a relationship between the stored reactive energy and the radiated field of each element, so that any change in phase inevitably alters its amplitude response to the incident field, unlike in conventional arrays, where amplitude and phase can be tuned independently~\cite{exp_dma1}. By controlling the resonance frequency of the metasurface elements, different forms of beam steering can be realized in DMA-based applications.}

\revised{Based on the reconfigurability of metasurfaces via their resonance frequencies}, in \cite{em_dma2} three methods were first introduced for achieving a single beam steering in a desired direction using a 1D metasurface array: Amplitude-Only Hologram (AOH), which controls variable amplitudes with constant phase near resonance frequency; Binary Amplitude Hologram (BAH), which toggles metasurfaces between on and off states; and the Lorentzian-Constrained Hologram (LCH)) which selects amplitudes [0 - 1] and phases ([0, $\pi$]) over the available region defined by a non-constant offset complex circle, namely, the Lorentzian circle. These three approaches were initially studied in \cite{DMA1} within a MIMO system employing DMA-assisted digital precoding, where the LCH demonstrated superior performance in beamforming and average sum-rate maximization over AOH and BAH. More recently, \cite{Yihan_Near} compared these methods in multiuser MISO communication networks with SINR guarantees, showing that LCH outperforms AOH and BAH in terms of channel gain, power efficiency, and beamforming. 

Being superior to AOH and BAH, LCH faces challenges in implementing beamforming algorithms for DMA-assisted architectures due to Lorentzian constraint, where the available weights are restricted by the Lorentzian-circle. Mapping-based solutions for projecting ideal unconstrained weights onto the Lorentzian-circle were first introduced in \cite{DMA3} for single beam-steering using a 1-D metasurface array. In \cite{DMA3}, the Generalized Method for LCH (GMLCH) was also implemented, exploring mappings from unitary modulus weights to the Lorentzian modulus circle, parameterized by the mapping center ($\alpha$) on the Lorentzian circle's imaginary axis. Three cases were compared based on different mapping centers ($\alpha$): Lorentzian-Constrained Phase Hologram (LCPH) ($\alpha=0$), Lorentzian-Constrained Euclidean Hologram (LCEH) ($\alpha=0.5$), and Lorentzian-Constrained Unitary Shift Hologram (LCUSH) ($\alpha=1.0$), corresponding to the bottom, center, and top points of the Lorentzian circle, \revised{respectively}. 
\revised{These mappings lead to different beamforming waveforms, particularly in terms of main-beam gain and grating-lobe behavior in single-beam setups. However, their impact on interference suppression in multiuser WIT systems remains an open question.}

Different resource allocation problems have been investigated for the joint optimization of digital precoders and DMA weights with a particular LCH method. In these works, LCUSH is the most commonly used mapping method \cite{DMA_BF1, DMA_BF2, DMA_app4, DMA_app5}, while LCEH has also been utilized in \cite{DMA1, Azarbahram_2024}.  In \cite{Azarbahram_2024}, Semi-Definite Programming (SDP) based on LCEH was employed to optimize DMA weights for beamforming in MISO-WPT systems, where LCEH is achieved by first solving the problem with a relaxed Lorentzian constraint and then projecting the obtained ideal weights onto the Lorentzian circle. On the other side, a common approach to obtain LCUSH involves first optimizing DMA weights on a unitary modulus circle using a manifold optimization technique such as Riemannian gradient, where the Lorentzian constraint is relaxed to phase-only weights with constant amplitude and arbitrary phase. Then, the ideal unconstrained weights are projected onto the Lorentzian circle by adding the imaginary term associated with the Lorentzian constraint (\( \alpha = 1.0 \)). This method has been applied in \cite{DMA_BF1, DMA_BF2, DMA_app5}, where manifold optimization with LCUSH was integrated into Alternating Optimization (AO) for weighted sum rate maximization in WIT networks. In comparison to these works, LCH involving the search for optimal weights on a unitary modulus circle has also been utilized in \cite{DMA_app3, DMA_app6, Yihan_Near}. However, instead of a separate projection step, the imaginary term associated with the Lorentzian constraint is explicitly involved in the cost function, yielding DMA weights in a closed-form manner. Particularly, in \cite{DMA_app3} AO based on manifold optimization was utilized for sum harvested power maximization problem. In \cite{DMA_app6}, Majorization-Minimization method was used for the optimization of DMA weights to achieve weighted sum rate maximization in a multi-user MISO-SWIPT network. In \cite{Yihan_Near}, successive convex approximation and alternating direction method of multipliers (ADMM) based AO algorithm was proposed to optimize digital precoders and DMA weights for minimizing total transmit power with SINR guarantees in XL-MIMO networks, comparing the performance of AOH and BAH with LCH but without addressing different mapping methods for LCH. More recently, \cite{DMA_Robert1} proposed a codebook design for a single-user MISO setup, highlighting the suboptimality of Lorentzian mapping for LCH systems by comparing LCPH and LCEH, but it did not address digital precoding or multi-user beamforming.

\revised{As highlighted in related works, different Lorentzian-mapping approaches are widely used in the literature, and the need for projection is not specific to SDP-based or manifold-based formulations. Rather, it stems from the physical tunability of DMA elements governed by Lorentzian resonance. While beamforming ideally requires weights of the form $e^{j\phi}$, the Lorentzian-constrained response inherently follows $\frac{j+e^{j\phi}}{2}$ structure. This structural mismatch requires a mapping or projection step in all Lorentzian-constrained DMA beamforming designs. To the best of our knowledge, this challenge has not been explicitly analyzed or systematically compared in wireless communication networks. Our work addresses this gap by introducing a unified and extensible framework that formalizes and improves the projection process.}  
\revised{Particularly, we consider a multi-user downlink system where the objective is to minimize total transmit power of DMA-aided setup while satisfying SINR constraints \cite{aaltinoklu}. Using GMLCH and an SDR-based algorithm, we jointly optimize the digital precoders and DMA weights, enabling different LCH types within a single framework, whereas prior works typically designed separate algorithms for each case. For example, \cite{Yihan_Near} proposed an SCA-ADMM method for the SINR guaranteed networks but did not allow comparisons across mapping strategies of LCH. Building on this foundation, we further propose a novel method called Adaptive Radius Lorentzian-Constrained Holography (ARLCH), which achieves enhanced optimization of DMA weights with better beamforming performance by leveraging additional degrees of freedom through the relaxation of the Lorentzian-circle diameter and dynamic adjustment of the mapping center within the optimization process. Our contributions can be summarized as following:}  


\begin{itemize} 
\item We develop a novel optimization algorithm based on SDP and AO for holographic beamforming using GMLCH in multi-user DMA-aided MISO\footnote{We use the term MISO since receiver users have a single antenna. However, DMA-aided MISO networks are often termed MIMO, as DMA act as HMIMO surfaces with MIMO-like two-dimensional precoding although a single antenna receiver is considered \cite{DMA_BF1, Yihan_Near}.} networks. The optimization equations are derived to optimize digital precoding vectors for fixed DMA weights, and to optimize relaxed DMA weights without Lorentzian constraints using Semi-Definite Relaxation (SDR) for fixed digital precoding vectors. The SDR solution provides an optimal boundary without Lorentzian constraints, which is then integrated with GMLCH to apply different LCH methods.
\item \revised{With the integration of GMLCH into convex optimization tools in the proposed algorithm, we obtain a unified platform to compare different LCH types (LCPH, LCEH, LCUSH) in the context of multiuser beamforming with QoS guarantees, a point not investigated systematically in the literature before.} Using this approach, we provide the comparisons for the performance gap depending on the parameters of GMLCH with respect to the optimal boundaries of DMA, and conventional FD architectures. Our results show that different mappings affect the performance gap, with LCEH outperforming LCUSH, and LCPH by providing better beamforming, especially for higher user numbers. 
\item We compare GMLCH against the ADMM-SCA-based baseline method in \cite{Yihan_Near}. Our results demonstrate that while the method in \cite{Yihan_Near} performs similarly to GMLCH (\(\alpha=1.0\), LCUSH), GMLCH (\(\alpha=0.5\), LCEH) achieves significantly better results. This highlights the robustness of our proposed optimization approach and highlights the critical role of selecting the appropriate mapping center in Lorentzian mapping.
\item We further demonstrate that relaxing the radius of the Lorentzian circle (i.e., the amplitude of the Lorentzian constraint) while preserving its form enables dynamic adjustment of the mapping center. This flexibility enhances beamforming capabilities in DMA-assisted precoding structures. Unlike previous works that constrain optimization within the unitary circle, we propose a novel approach, ARLCH, which significantly improves performance in SINR-guaranteed networks for randomly located multi-user scenarios. \revised{We used GMLCH as benchmarking tool for our novel ARLCH approach in order to show performance gain against different schemes available in literature}. Furthermore, we conduct Monte-Carlo simulations for multi-user setups. The results show that ARLCH outperforms the best GMLCH method (LCEH), with its advantage growing as the number of users grows.
\end{itemize}
\revised{Building upon the same SDP formulation for obtaining unrestricted DMA weights, this paper significantly extends our preliminary work in \cite{aaltinoklu} by: (i) integrating the GMLCH into an alternating optimization framework for joint design of digital precoders and DMA weights; (ii) proposing a novel adaptive mapping algorithm, ARLCH, which enhances Lorentzian mapping and beamforming performance; and (iii) presenting extensive simulation results that analyze the impact of various Lorentzian-constrained weighting schemes, LCEH, LCPH, and LCUSH, none of which were studied in \cite{aaltinoklu} or, to the best of our knowledge, in prior literature.}

The remainder of the paper is organized as follows: Section~II presents the system model, Section~III the optimization problems via SDP, Section~IV the GMLCH and proposed ARLCH schemes, while Sections~V and~VI provide discussions and numerical results, respectively.

\noindent\textit{Notation:} 
A matrix is denoted by the boldface capital \(\mathbf{W}\) with \(\text{rank}(\mathbf{W})\), \(\text{Tr}(\mathbf{W})\), \(\mathbf{W}^T\), \(\mathbf{W}^H\), and \(\text{Vec}(\mathbf{W})\) representing its rank, trace, transpose, Hermitian conjugate, and vectorization, respectively. A vector is denoted by the boldface lower-case \(\mathbf{w}\), with \(\|\mathbf{w}\|\), \(\mathbf{w}^T\), and \(\mathbf{w}^H\) indicating its Euclidean norm, transpose, and Hermitian conjugate, respectively. The unit vector in the direction of \(\mathbf{w}\) is denoted by \(\hat{\mathbf{w}}\), while the scalar is represented by \(w\), and \(|w|\) is its absolute value. \revised{The operator \(\operatorname{Re}(\cdot)\) denotes the real part of a complex number. The Kronecker product is indicated by \(\otimes\), and the symbol \(\circ\) denotes Hadamard product.}

\begin{figure}[!t]
\centering
\includegraphics[width=0.4\textwidth, trim=0.1in 0.1in 0.8in 0.0in, clip]{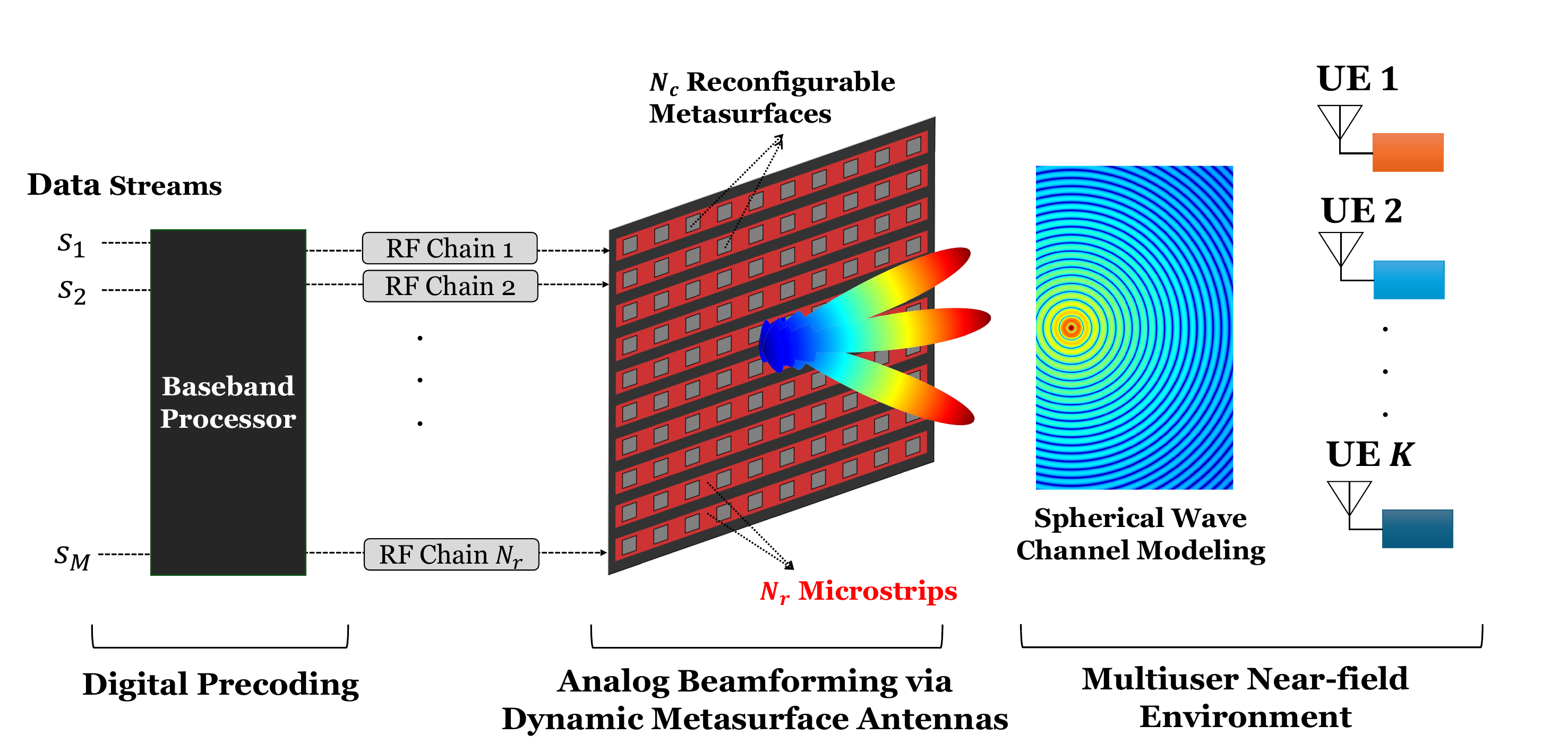}
\caption{\revised{DMA aided Multi-user downlink MISO system}.}
\label{fig:system_setup}
\end{figure}

\section{SYSTEM MODEL \& PROBLEM FORMULATION}
We consider a multi-user downlink MISO system, where the base station (BS) is equipped with DMA-aided beamforming architecture, as shown in Fig. \ref{fig:system_setup}. In addition, benchmark solutions are provided for the scenario in which the BS utilizes an FD architecture. In this system, the BS serves \(K\) users, each equipped with a single antenna and requesting a certain level of SINR, in a generic Line-of-Sight (LoS) spherical wave channel model that covers near-field users. Moreover, perfect knowledge of the channel is assumed.

\subsection{Signal Processing for DMA Architecture}
The DMA-based architecture comprises of \(N_\text{r}\) microstrips each containing \(N_\text{c}\) metasurface-based intrinsic elements, yielding \(N \triangleq N_\text{c} N_\text{r}\) of total elements. In this setup, each microstrip is connected to the digital beamformer through a single RF chain and the complex amplitudes of the reference wave for each microstrip are controlled by digital beamforming vectors \(\mathbf{w}_m \in \mathbb{C}^{N_\text{r} \times 1}\) for all symbols \revised{$s_m$ with} \(m \in M\), where \(M = \min(K, N_\text{r})\). The radiated pattern of individual elements is further adjusted with their corresponding dynamically configurable weights \revised{(\(\mathbf{Q} \in \mathbb{C}^{N \times N_\text{r}}\))}. Hence, the transmitted signal for DMA is
\begin{equation}
\mathbf{x} = \sum_{m=1}^{M} \mathbf{x}_m = \sum_{m=1}^{M} \mathbf{H} \mathbf{Q} \mathbf{w}_m s_m,
\label{eq:6}
\end{equation}
where, \(\mathbf{H}\) is a diagonal matrix with elements \(\mathbf{H}_{(i-1)N_\text{c} + l, (i-1)N_\text{c} + l} = e^{-d_{i,l} (\alpha_i + j \beta_i)}\), where \(d_{i,l}\) is the position of the \(l\)-th element along the \(i\)-th microstrip, and \(\alpha_i, \beta_i\) are its attenuation and propagation constants \cite{metasurface_6g}. Matrix \(\mathbf{Q}\) includes the inter-connectivity of the individual elements to the excitation ports, with their frequency-dependent response to the external excitation, i.e., the Lorentzian resonance response, 
\begin{equation}
\label{eq:7p}
\mathbb{Q} = \left\{ q = \frac{j + e^{j\Phi}}{2} : \Phi \in [0, 2\pi] \right\}.
\end{equation} 
In \eqref{eq:7p}, the amplitudes and phases of the individual elements \( q\) are dependent on each other, and the phases are limited to the range \([0, \pi]\). 
Based on this, \(\mathbf{Q}\) can be defined in a block-diagonal form, as described in \cite{DMA_BF1} with
\begin{equation}
\label{eq:7}
\mathbf{Q}_{(i-1)N_\text{c} + l, n} =
\begin{cases}
q_{i,l} \in \mathbb{Q}  & \text{if } i = n \\
0 & \text{if } i \ne n.
\end{cases}
\end{equation}
Furthermore, we use the near-field channel model based on spherical wave propagation given in \cite{DMA_BF1} with the path-loss modelling defined in \cite{gain_element}. Accordingly, the radiation pattern of the elements is well-approximated with:
\begin{equation}
\label{eq:1}
G_\text{e}(\psi) = 
\begin{cases} 
2(g+1) \cos^{g}(\psi), & 0 \leq \psi \leq \pi/2 \\
0, & \pi/2 < \psi \leq \pi 
\end{cases}
\end{equation}
where $\psi$ is the angle measured from the Boresight of the array, \(g\) is the measure of antenna gain. 
As the elements with \(G_\text{e}(\psi)\) radiate over the free-space, propagation of electromagnetic waves towards user position ($\mathbf{r}_k$) from the ($i$,$l$)-th element of the uniform planar array (UPA) (\(\mathbf{r}_{i,l}\)) can be expressed for the near-field channel as follows:
\begin{equation}
\label{eq:2}
\gamma_{k}(i, l) = \sqrt{G_\text{e}(\psi)} \frac{\lambda}{4\pi \|\mathbf{r}_k - \mathbf{r}_{i, l}\|} e^{-j \beta_0 \|\mathbf{r}_k - \mathbf{r}_{i, l}\|},
\end{equation}
where \(\lambda\), and \(\beta_0\) are the free-space wavelength and wavenumber, respectively. The entries \(\gamma_{k}(i, l)\) are the elements of the channel vector \(\boldsymbol{\gamma}_k \in \mathbb{C}^{N \times 1}\), i.e., \(\boldsymbol{\gamma}_k \triangleq \begin{bmatrix}
\gamma_{k}(1, 1), \gamma_{k}(1, 2), \ldots, \gamma_{k}(N_\text{r}, N_\text{c})
\end{bmatrix}^H\).

Finally, by inserting the antenna aperture distributions from (\ref{eq:6}) into the channel model given in (\ref{eq:2}), the received signal at User \(k\) can be formulated as
\begin{equation}
\label{eq:8}
{y}_{k} = \boldsymbol{\gamma}_k^H \sum_{m=1}^{M} \mathbf{H} \mathbf{Q} \mathbf{w}_m s_m + n_k,
\end{equation}
\revised{where $n_k \sim \mathcal{CN}(0,\sigma_k^2)$ denotes the additive white Gaussian noise (AWGN) at User $k$.}
\subsection{Problem Formulation}
Given that optimization aim of the downlink beamforming problem is to minimize the total transmit power under the constraints of given sets of SINR for different users, and with \revised{\( \mathbf{x}_m = \mathbf{H} \mathbf{Q} \mathbf{w}_m s_m\) and assuming unit-power symbols $\mathbb{E}[|s_m|^2] = 1$ for all $m$, the problem can be expressed as}
\begin{subequations}
\label{eq:12}
\begin{align}
& \underset{\mathbf{Q}, \mathbf{w}_m, \forall m}{\text{minimize}} \quad \sum_{m=1}^{M} \|\mathbf{H} \mathbf{Q} \mathbf{w}_m\|^2 \label{eq:12a}\\
\text{s.t.} & \quad \frac{|\boldsymbol{\gamma}_k^H \mathbf{H} \mathbf{Q} \mathbf{w}_k|^2}{\sum_{\substack{m=1 \\ m \neq k}}^{M} |\boldsymbol{\gamma}_k^H \mathbf{H} \mathbf{Q} \mathbf{w}_m|^2 + \sigma_k^2} \geq \delta_k, \quad \forall k, \label{eq:12b}\\
& \quad q_{n} \in \mathbb{Q}, \quad \forall n, \label{eq:12c}
\end{align}
\end{subequations}
where \(\delta_1, \dots, \delta_k\) represent the SINR thresholds that must be guaranteed for each user, and \(\sigma_k^2\) is the noise power for the \(k\)-th user. In \eqref{eq:12}, \(q_{n}\) is the DMA weight of the \(l\)-th element along the \(i\)-th microstrip, given by \(n=(i-1)N_c + l \), and \(n \in N \). For FD-based architecture, the problem in \eqref{eq:12} can be simplified with \( \mathbf{x}_m = \mathbf{w}_m\).  The global optimum solutions for the digital precoding vectors \( \mathbf{w}_m \) can be obtained by solving the tractable form of the problem via SDR \cite{SDP}. 

\section{BEAMFORMING OPTIMIZATION SOLUTION FOR DMA}


The solution of problem~\eqref{eq:12} requires joint design of the digital precoders and the DMA weights. \revised{Since constraint~\eqref{eq:12c} is not tractable and difficult to handle directly, we adopt an alternating optimization approach in a more tractable form using convex optimization tools by relaxing this constraint. Specifically, we decouple the problem into two SDR subproblems: one for the digital precoders (\(\mathbf{w}_m\)) and one for the DMA weights (\(\mathbf{Q}\)), both derived from a relaxed form of~\eqref{eq:12}. The Lorentzian constraint~\eqref{eq:12c} is then enforced externally through a projection step within each iteration, as detailed in Section~IV. The resulting SDR formulations are solved using the CVX convex optimization toolbox~\cite{cvx1}.}
\subsection{Substage 1: Optimizing the Digital Precoder}
When \(\mathbf{Q}\) is fixed, the total transmitted power \(P_{\text{Tx}}\) can be derived as follows:
\begin{equation}
\label{eq:13}
\begin{aligned}
P_{\text{Tx}} &= \sum_{m=1}^{M} \text{Tr}(\mathbf{X}_m) = \sum_{m=1}^{M} \text{Tr} \left(\mathbf{H} \mathbf{Q} \mathbf{w}_m (\mathbf{H} \mathbf{Q} \mathbf{w}_m)^H\right) \\
&= \sum_{m=1}^{M} \text{Tr}(\mathbf{Z} \mathbf{W}_m),
\end{aligned}
\end{equation}
where \(\mathbf{Z} = (\mathbf{H} \mathbf{Q})^H \mathbf{H} \mathbf{Q}\).
Moreover, the received power at User \(k\) due to the \(m\)-th beamforming vector of the DMA \((P_{\text{Rx},k,m})\) can be derived as a function of \(\mathbf{W}_m\) according to
\begin{equation}  
\label{eq:14}
\begin{aligned}
P_{\text{Rx},k,m} &= \text{Tr}(\boldsymbol{\gamma}_k^H \mathbf{H} \mathbf{Q} \mathbf{w}_m \mathbf{w}_m^H (\boldsymbol{\gamma}_k^H \mathbf{H} \mathbf{Q})^H) \\ &= \text{Tr}(\mathbf{P}_k \mathbf{W}_m),
\end{aligned}
\end{equation}  
where \revised{\(\mathbf{P}_k = (\boldsymbol{\gamma}_k^H \mathbf{H} \mathbf{Q})^H\boldsymbol{\gamma}_k^H \mathbf{H} \mathbf{Q} \)} and \revised{\(\mathbf{W}_m = \mathbf{w}_m \mathbf{w}_m^H\)}. Then, combining \eqref{eq:13} and \eqref{eq:14}, optimization problem \eqref{eq:12} can be reformulated in SDP relaxation form as
\begin{equation}
\label{eq:15}
\begin{aligned}
 \underset{\mathbf{W}_m}{\text{minimize}} \quad &\sum_{m=1}^{M} \text{Tr}(\mathbf{Z} \mathbf{W}_m)  \\
\text{s.t.}  \quad \text{Tr}(\mathbf{P}_k \mathbf{W}_k) - \delta_k &\sum_{\substack{m=1 \\ m \neq k}}^{M} \text{Tr}(\mathbf{P}_k \mathbf{W}_m) - \delta_k \sigma_k^2 \geq 0, \quad \forall k, \\
& \quad \mathbf{W}_m \succeq 0, \, \quad \forall m.
\end{aligned}
\end{equation}

After solving the SDP problem defined in \eqref{eq:15}, the digital precoding vectors \(\mathbf{w}_m \in \mathbb{C}^{N_\text{r} \times 1}\) can be obtained via eigenvalue decomposition of the corresponding matrix \(\mathbf{W}_m\). \revised{This solution is globally optimal, as the optimal \(\mathbf{W}_m\) is guaranteed to be rank-one under the considered formulation, as shown in \cite{SDP}.}

\subsection{Substage 2: Optimizing the DMA Weights}

\label{subsec:optimizing_dma_weights}

When \(\mathbf{w}_m\) for \(\forall m\) is fixed,  utilizing the identity \(\mathbf{A}^T \mathbf{Q} \mathbf{b} = \left(\mathbf{b}^T \otimes \mathbf{A}^T\right) \text{Vec}(\mathbf{Q})\) \cite{DMA_BF1}, \(\mathbf{x}_m\) can be rewritten as:
\begin{equation}
\label{eq:16}
\mathbf{x}_m = \mathbf{H} \mathbf{Q} \mathbf{w}_m = \left(\mathbf{w}_m^T \otimes \mathbf{H}\right) \text{vec}(\mathbf{Q}).
\end{equation}
Then, by defining \(\mathbf{A}_m = \left(\mathbf{w}_m^T \otimes \mathbf{H}\right)^H \in \mathbb{C}^{L \times N}\) and the vector \(\mathbf{q} = \text{vec}(\mathbf{Q}) \in \mathbb{C}^{L \times 1}\), where \(L = N_\text{r}^2 N_\text{c}\), the total transmitted power \(P_{\text{Tx}}\) can be derived as follows:
\begin{equation}
\begin{aligned}
\label{eq:17}
P_{\text{Tx}} = \sum_{m=1}^{M} \text{Tr}\left(\mathbf{X}_m\right) &= \sum_{m=1}^{M} \text{Tr}\left( \mathbf{A}_m^H \mathbf{q} \mathbf{q}^H \mathbf{A}_m \right).
\end{aligned}
\end{equation}
By defining \(\mathbf{\tilde{q}} \in \mathbb{C}^{N \times 1}\), obtained by removing all the zero elements from \(\mathbf{q}\), and \(\mathbf{\tilde{A}}_m \in \mathbb{C}^{N \times N}\), formed by removing the rows corresponding to the indices of the removed elements in \(\mathbf{q}\), \eqref{eq:17} can be rewritten as:
\begin{equation}
\begin{aligned}
\label{eq:18}
P_{\text{Tx}} = \sum_{m=1}^{M} \text{Tr}\left( \mathbf{\tilde{A}}_m^H \mathbf{\tilde{q}} \mathbf{\tilde{q}}^H \mathbf{\tilde{A}}_m \right) &= \sum_{m=1}^{M} \text{Tr}\left( \mathbf{\tilde{B}}_m \mathbf{\tilde{Q}} \right),
\end{aligned}
\end{equation}
where \(\mathbf{\tilde{B}}_m = \mathbf{\tilde{A}}_m \mathbf{\tilde{A}}_m^H\) and \(\mathbf{\tilde{Q}} = \mathbf{\tilde{q}} \mathbf{\tilde{q}}^H\).

Following the derivations for \(P_{\text{Tx}}\), 
the received power at User \(k\) due to the \(m\)-th beamforming vector of the DMA \(P_{\text{Rx},k,m}\) can be derived similarly.  Given the fact that \(\mathbf{a}^T \mathbf{Q} \mathbf{b} = \left(\mathbf{b}^T \otimes \mathbf{a}^T\right) \text{vec}(\mathbf{Q})\), \( y_{k,m} \) given in \eqref{eq:8} can be defined and reformulated as
\begin{equation}
\begin{aligned}
\label{eq:19}
 = \mathbf{c}_{k,m}^H \mathbf{q},
\end{aligned}
\end{equation}
where \(\mathbf{c}_{k,m} = \left(\mathbf{w}_m^T \otimes (\boldsymbol{\gamma}_k^H \mathbf{H})\right)^H \in \mathbb{C}^{L \times 1}\). Furthermore, the modified vectors, obtained by removing zero elements as described above, can be defined as \(\mathbf{\tilde{c}}_{k,m} \in \mathbb{C}^{N \times 1}\) and \(\mathbf{\tilde{q}} \in \mathbb{C}^{N \times 1}\). Then,  \(P_{\text{Rx},k,m}\) can be further simplified as:
\begin{equation}
\label{eq:20}
\begin{aligned}
P_{\text{Rx},k,m} &= \text{Tr}\left(\mathbf{\tilde{c}}_{k,m}^H \mathbf{\tilde{q}}(\mathbf{\tilde{c}}_{k,m}^H \mathbf{\tilde{q}})^H\right)= \text{Tr}\left(\mathbf{\tilde{C}}_{k,m} \mathbf{\tilde{Q}}\right),
\end{aligned}
\end{equation}
where \(\mathbf{\tilde{C}}_{k,m} = \mathbf{\tilde{c}}_{k,m} \mathbf{\tilde{c}}_{k,m}^H\).
\vspace{0.01in} 

Finally, by combining \eqref{eq:18} and  \eqref{eq:20}, the optimization problem is formulated as 
\begin{equation}
\label{eq:21}
\begin{aligned}
 \underset{\mathbf{\tilde{Q}}}{\text{minimize}} \quad &\sum_{m=1}^{M} \text{Tr}(\mathbf{\tilde{B}}_m \mathbf{\tilde{Q}}) \\
\text{s.t.}  \quad \text{Tr}(\mathbf{\tilde{C}}_{k,k} \mathbf{\tilde{Q}}) - \delta_k &\sum_{\substack{m=1 \\ m \neq k}}^{M} \text{Tr}(\mathbf{\tilde{C}}_{k,m} \mathbf{\tilde{Q}}) - \delta_k \sigma_k^2 \geq 0, \quad \forall k, \\
& \quad \mathbf{\tilde{Q}} \succeq 0.
\end{aligned}
\end{equation}

For given digital precoding vectors \(\{\mathbf{w}_m^\star\} \, \forall m \in M\), the solution of \eqref{eq:21} yields the optimal matrix \(\mathbf{\tilde{Q}}^\star \in \mathbb{C}^{N \times N}\). \revised{The problem in \eqref{eq:21} is relaxed by omitting the rank-one constraint. To recover the beamforming vector \(\mathbf{\tilde{q}}^\star \in \mathbb{C}^{N \times 1}\), we employ a best rank-one approximation using the eigenvalue decomposition of \(\mathbf{\tilde{Q}}^\star\), extracting the dominant eigenvalue–eigenvector pair, as detailed in \cite{5447068}.}

Next, the solution vector \(\mathbf{\tilde{q}}^\star\) must be mapped onto the Lorentzian circle in \eqref{eq:7p} as represented by
\begin{equation}
\label{eq:22}
\mathbf{\tilde{q}}^\star \in \mathbb{C}^{N \times 1} \longrightarrow \mathbf{\tilde{q}} \in \mathbb{Q}^{N \times 1}.
\end{equation} 

\revised{For this mapping step, we employ various forms of Lorentzian mapping, which are presented in detail in Section~\ref{sec:LORENTZIAN_MAPPING_FOR_ANALOG_BEAMFORMING}. 
The overall solution framework involves solving the individual problems \eqref{eq:15} and \eqref{eq:21} in an alternating manner: the digital precoding vectors are iteratively updated toward optimality, while the DMA weights are optimized through constrained projection using the selected Lorentzian mapping. This alternating optimization procedure is summarized in Algorithm~\ref{alg:proposed_algorithm}, which integrates both the precoder updates and Lorentzian-constrained weight mappings introduced in this work. The optimization procedure outlined in Algorithm~\ref{alg:proposed_algorithm}, along with the Lorentzian mapping step, ensures that the resulting DMA weights adhere to the Lorentzian resonance response introduced in \eqref{eq:7p}. This resonance model imposes a sinusoidal variation in the amplitudes of the DMA weights as a function of their phases, whereas beamforming typically requires phase alignment, as assumed in the channel model. Therefore, the convergence and effectiveness of Algorithm~\ref{alg:proposed_algorithm} critically depend on the method used to project from ideal phase alignment to the Lorentzian-constrained domain with amplitude variation. The impact of this projection and different mapping strategies will be explored in the following section.
}

\begin{algorithm}[t]
\caption{Proposed algorithm for solving problem \eqref{eq:12}}
\label{alg:proposed_algorithm}
\begin{algorithmic}[1]
\State \textbf{Initialize:} $\mathbf{Q}^{(0)}$;
\State Solve \eqref{eq:15} to calculate $\left\{\mathbf{W}^{(0)}\right\}_{m=1}^{M}$;
\State Update $\left\{\mathbf{w}^{(0)}\right\}_{m=1}^{M}$ and $P_{\text{Tx}}^{(0)}$ based on $\left\{\mathbf{W}^{(0)}\right\}_{m=1}^{M}$;
\For{$t = 1, \dots, T$}
    \State Solve \eqref{eq:21} to calculate \(\mathbf{\tilde{Q}}^\star \in \mathbb{C}^{N \times N}\) based on $\left\{\mathbf{w}^{(t-1)}\right\}_{m=1}^{M}$ ;
    \State Calculate $\mathbf{\tilde{q}}^\star \in \mathbb{C}^{N \times 1}$ based on \(\mathbf{\tilde{Q}}^\star \in \mathbb{C}^{N \times N}\);
    \State Calculate $\mathbf{\tilde{q}}$ with Lorentzian mapping (\eqref{eq:22}) of $\mathbf{\tilde{q}}^\star \in \mathbb{C}^{N \times 1}$;
    \State Update $\mathbf{Q}^{(t)}$ for problem \eqref{eq:12} based on $\mathbf{\tilde{q}}$ and \eqref{eq:7};
    \State Solve \eqref{eq:15} to calculate $\left\{\mathbf{W}^{(t)}\right\}_{m=1}^{M}$ based on $\mathbf{Q}^{(t)}$;
    \State Update $\left\{\mathbf{w}^{(t)}\right\}_{m=1}^{M}$ and $P_{\text{Tx}}^{(t)}$ based on $\left\{\mathbf{W}^{(t)}\right\}_{m=1}^{M}$;

\If {$P_{\text{Tx}}^{(t)} \leq P_{\text{Tx}}^{(t-1)}$}
    \State $\left\{\mathbf{w}^{(f)}\right\}_{m=1}^{M} \leftarrow \left\{\mathbf{w}^{(t)}\right\}_{m=1}^{M}$; $\mathbf{Q}^{(f)} \leftarrow \mathbf{Q}^{(t)}$;
    \State $P_{\text{Tx}}^{(f)} \leftarrow P_{\text{Tx}}^{(t)}$;
\EndIf
\EndFor
\State \textbf{Output:} $\left\{\mathbf{w}^{(f)}\right\}_{m=1}^{M}$, $\mathbf{Q}^{(f)}$, $P_{\text{Tx}}^{(f)}$.
\end{algorithmic}
\end{algorithm}

\revised{
It should be noted that during Step~7 of Algorithm~\ref{alg:proposed_algorithm}, the SINR constraints previously satisfied in Step~5 may be temporarily violated due to the Lorentzian mapping. This is immediately addressed in Step~9, where Problem~\eqref{eq:15} is solved to re-optimize the digital precoder set \(\{\mathbf{w}_m\}\) for the updated Lorentzian-constrained weights, thereby fully restoring the SINR constraints. Consequently, the final output of each iteration of Algorithm~\ref{alg:proposed_algorithm} yields Lorentzian-constrained DMA weights and corresponding digital precoding vectors that jointly satisfy the SINR requirements. Finally, the alternating execution of Algorithm~\ref{alg:proposed_algorithm} effectively drives the convergence of digital precoder vectors toward their optimal solution, while simultaneously optimizing the DMA weights toward high-quality suboptimal solutions supported by the proposed mapping schemes.
}

\section{LORENTZIAN MAPPING FOR ANALOG BEAMFORMING}

\label{sec:LORENTZIAN_MAPPING_FOR_ANALOG_BEAMFORMING}

In this section\footnote{For simplicity, the tilde notation is omitted throughout this section.}, we present the various methods for Lorentzian \revised{m}apping, as represented in \eqref{eq:22}, to obtain the optimal Lorentzian-constrained points \(\mathbf{q} \in \mathbb{Q}^{N \times 1}\) for the DMA weights from ideal weights obtained via the solution of SDR problem \(\mathbf{q}^\star \in \mathbb{C}^{N \times 1}\).

During the conduction of Algorithm \ref{alg:proposed_algorithm}, Step 6 solves the SDR problem, yielding a solution in the form of:
\begin{equation}
\label{eq:23}
 q_n^\star = |q_n^\star| e^{j\phi_n^\star}, \quad \forall n \in N,
\end{equation}
\revised{where $|q_n^\star| \in \mathbb{R}$ and $\phi_n^\star \in [0, 2\pi]$ denote the amplitude and phase of the $n$-th element of the solution vector $\mathbf{q}^\star$, respectively}. Since the SDR approach relaxes the Lorentzian constraint, \(\mathbf{q}^\star\) is not strictly confined to a constant, unitary, and Lorentzian circular manifold (i.e, Lorentzian circle). \revised{We refer to these weights as ideal or unrestricted. These represent optimal phase alignments dictated by user spatial locations and the underlying channel model. When Algorithm~\ref{alg:proposed_algorithm} is executed without enforcing Lorentzian mapping (i.e., omitting Step 7), the solution closely approximates the unconstrained performance boundary. However, this unconstrained solution does not reflect the physical tunability limitations of DMAs, which are governed by Lorentzian resonance behavior as defined in \eqref{eq:7p}. To address this, the final step of our algorithm includes a Lorentzian mapping process that projects the ideal weights onto a Lorentzian-constrained circle. This projection step is common in DMA beamforming applications~\cite{DMA_BF1, DMA_BF2, DMA_app4, DMA_app5, Azarbahram_2024}, although each prior study adopts different projection schemes and optimization strategies to derive the ideal (unrestricted) weights}. 

\revised{In this section, we first present a unified projection framework, named GMLCH, which standardizes this mapping process onto the Lorentzian circle for a fixed mapping center. This framework enables comparative evaluation of several projection strategies used in the literature, such as LCUSH, LCPH, and LCEH, as previously discussed in detail in the Introduction section. Subsequently, we propose a novel adaptive Lorentzian mapping method, i.e., ARLCH, which improves beamforming performance by minimizing the discrepancy between ideal and physically feasible weights through dynamic phase optimization on a relaxed Lorentzian circle with an adaptive radius, while strictly preserving the unitary amplitude of Lorentzian constraint described in \eqref{eq:7p}.}

\subsection{Generalized Method for Lorentzian-Constrained Holography}
\begin{figure*}[!t]
\centering
\subfloat[]{\includegraphics[trim={0.0in 0.0in 0.0in 0.0in},width=0.24\textwidth]{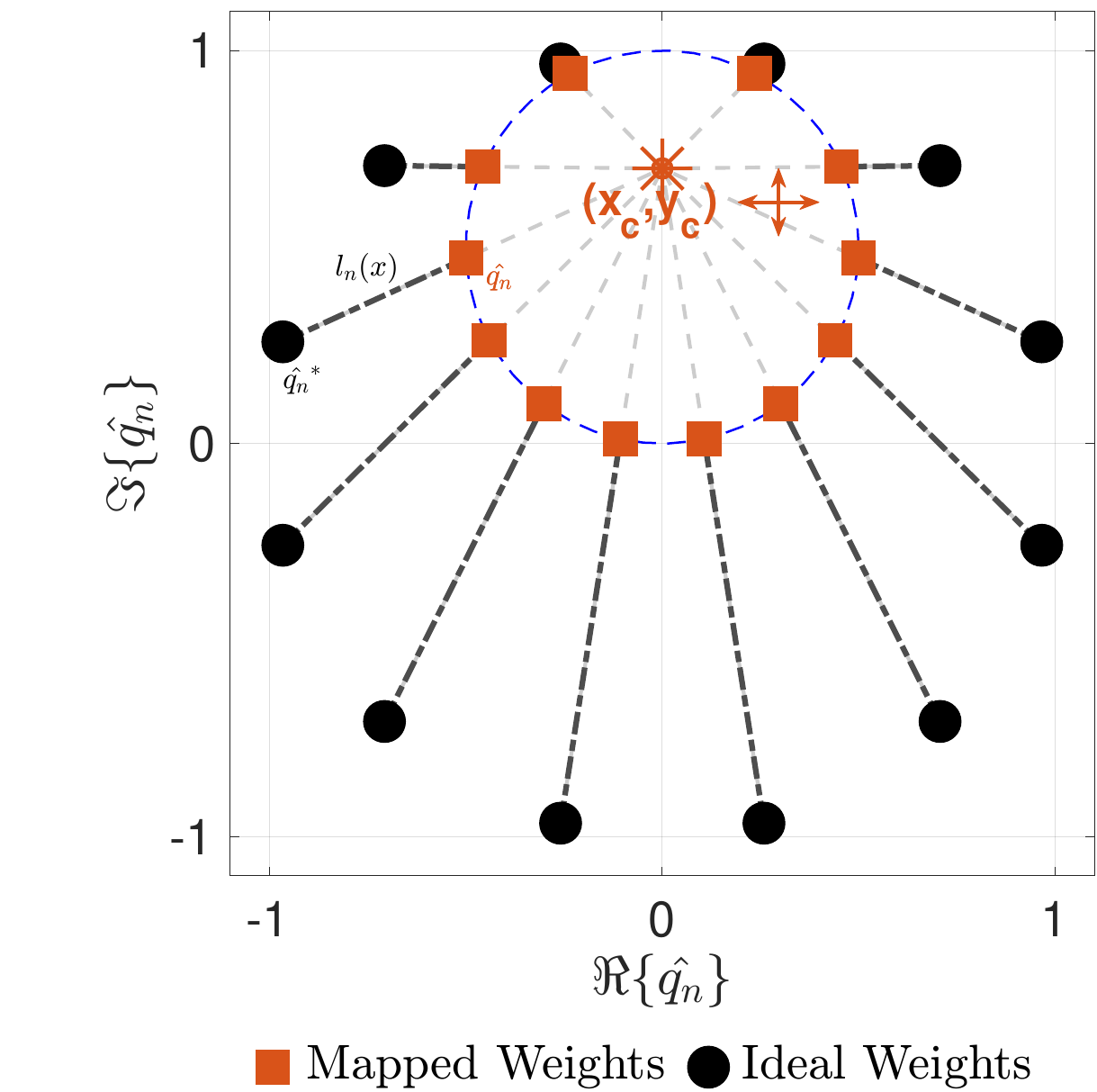}%
\label{fig:GMLCH_fig1}}
\hfil
\centering
\subfloat[]{\includegraphics[trim={0.0in 0.0in 0.0in 0.0in},width=0.24\textwidth]{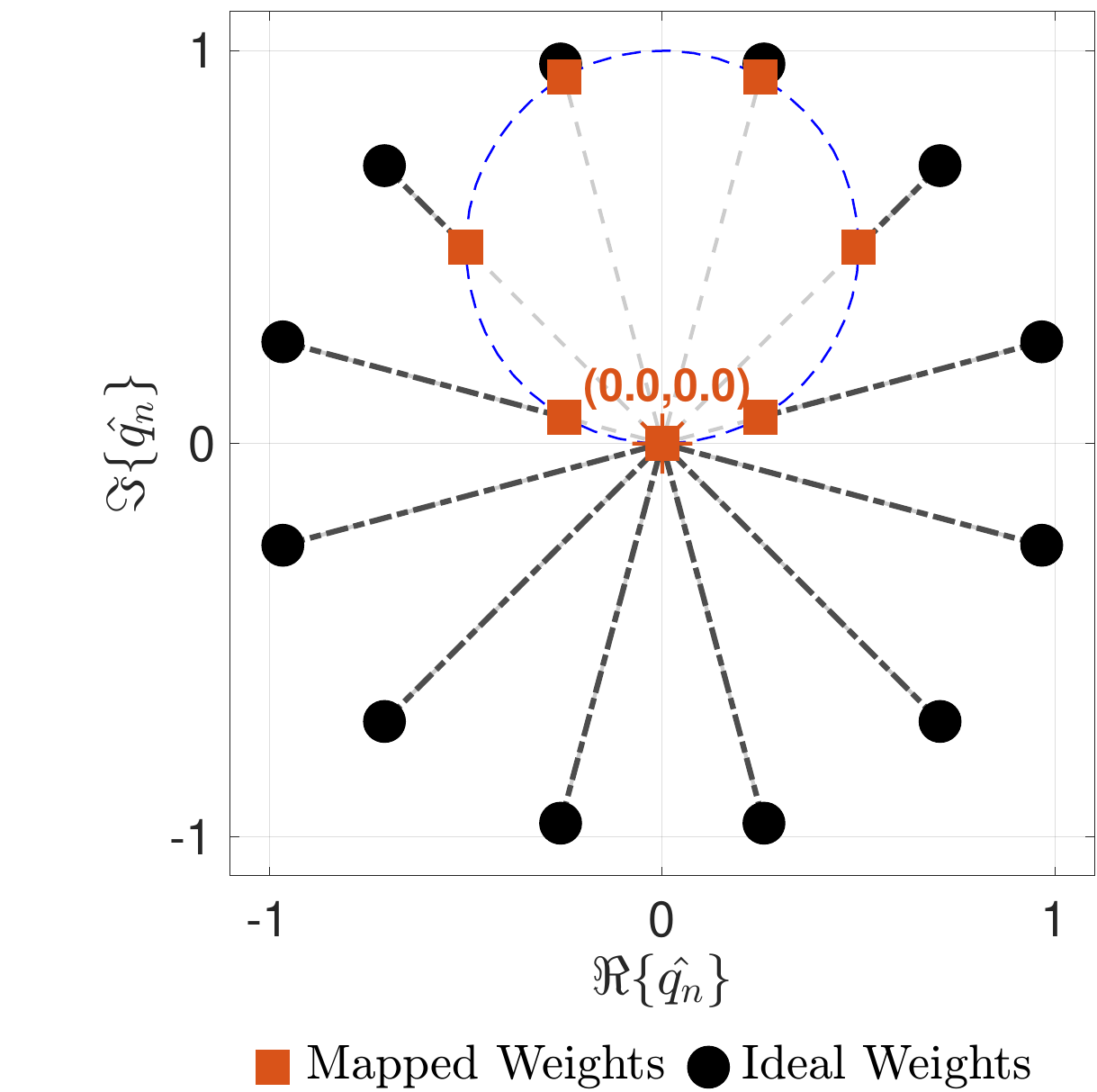}%
\label{fig:GMLCH_sub1}}
\hfil
\subfloat[]{\includegraphics[trim={0.0in 0.0in 0.0in 0.0in},width=0.24\textwidth]{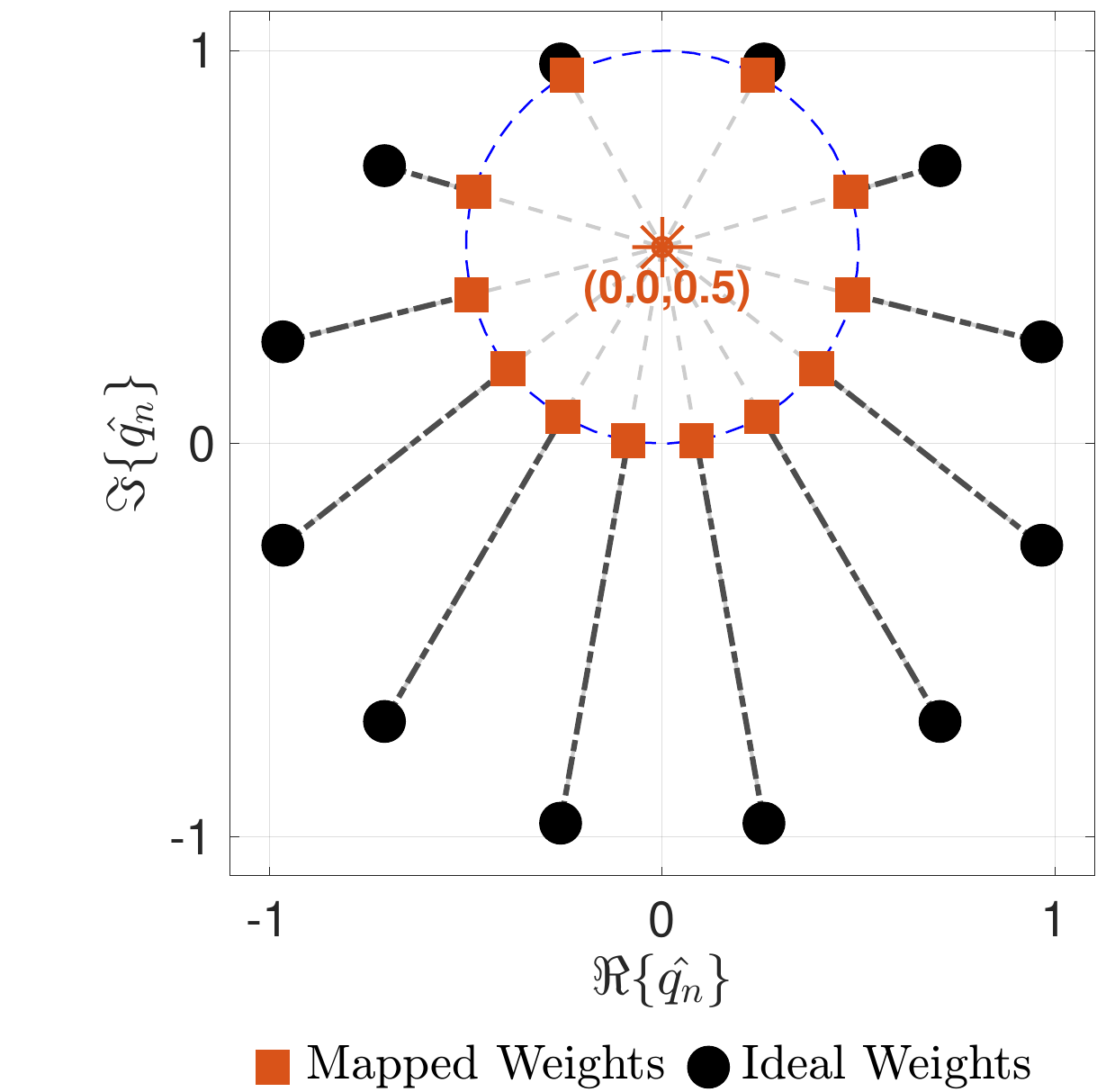}
\label{fig:GMLCH_sub2}}
\subfloat[]{\includegraphics[trim={0.0in 0.0in 0.0in 0.0in},width=0.24\textwidth]{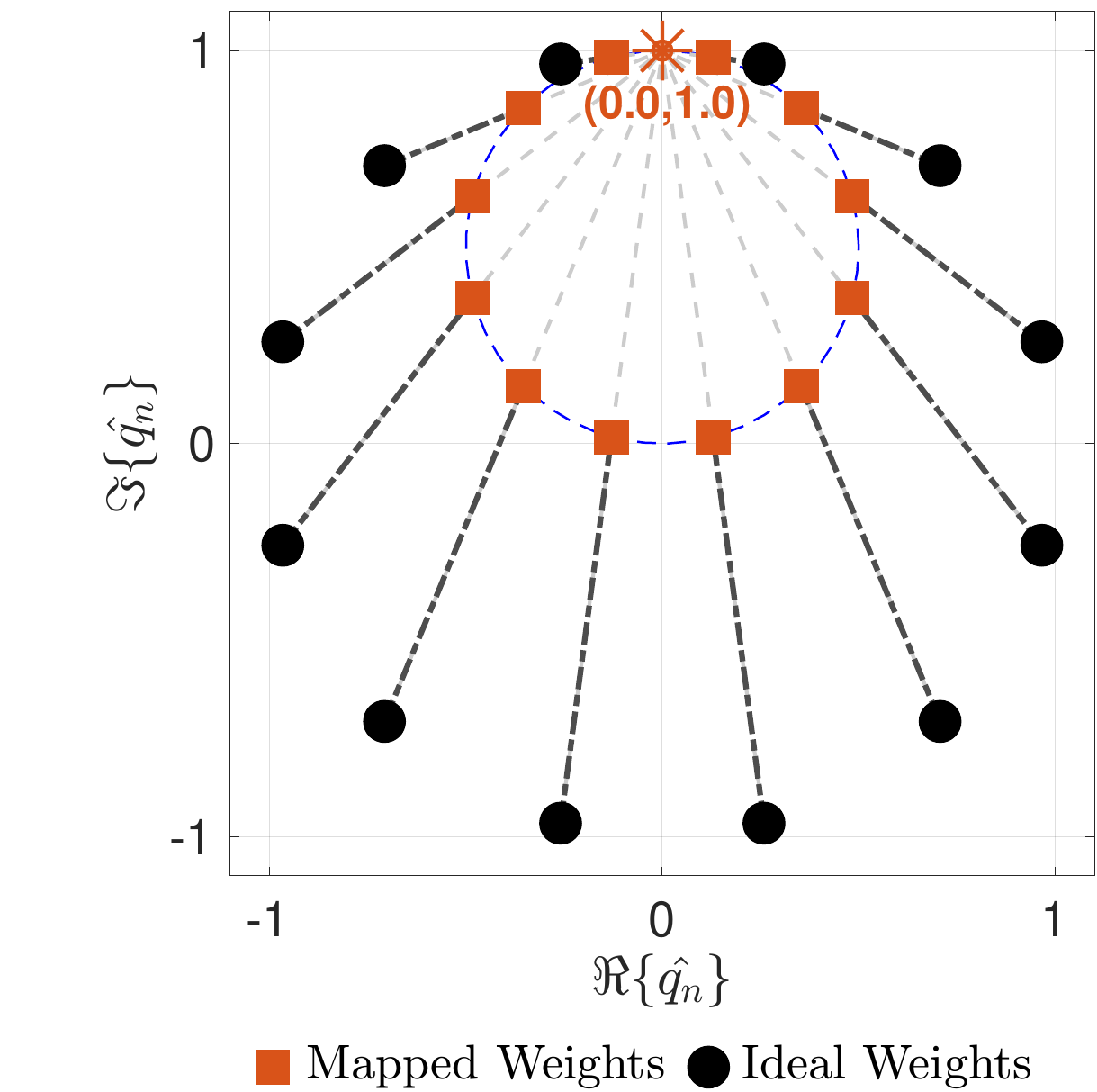}
\label{fig:GMLCH_sub3}}

\caption{Mapping of ideal DMA weights onto the Lorentzian circle (\(r=0.5\)) via GMLCH, parameterized by \((x_\text{c}, y_\text{c})\). (a) GMLCH with any \((x_\text{c}, y_\text{c})\) (b) LCPH with \((x_\text{c}, y_\text{c}) = (0.0, 0.0)\), (c) LCEH with \((x_\text{c}, y_\text{c}) = (0.0, 0.5)\), and (d) LCUSH with \((x_\text{c}, y_\text{c}) = (0.0, 1.0)\).}

\label{fig:GMLCH_fig_ex}
\end{figure*}

GMLCH projects ideal weights onto a Lorentzian circle of unitary amplitude \((r = 0.5)\), parameterized by its center in real and imaginary axis with \((x_\text{c}, y_\text{c})\), as shown in Fig.~\ref{fig:GMLCH_fig1}.
This mapping is applied to the unitary form of the solution vector \(\mathbf{q}^\star\) in \eqref{eq:23}, representing the ideal weights of individual DMA elements:
\begin{equation}
\label{eq:25}
\hat{q}_n^\star 
=  e^{j \phi_n^\star}
= \cos(\phi_n^\star) + j\,\sin(\phi_n^\star),
\quad \forall n \in N,
\end{equation}
where \(\phi_n^\star\) is the corresponding phase for \(n\)-th element.

As presented in Fig. \ref{fig:GMLCH_fig1}, GMLCH can be defined as finding the intersection points between the Lorentzian-constrained circle and lines drawn from the ideal weights \(\mathbf{\hat{q}}_n^\star\) to the pre-defined center \((x_\text{c}, y_\text{c})\). Based on the given expression in \eqref{eq:25}, the equation of the line passing through the ideal weight of the \(n\)-th element \({\hat{q}}_n^\star\) and \((x_\text{c}, y_\text{c})\) can be defined by:
\begin{equation}
\label{eq:26}
l_n(x)
= x + j \left[
    \frac{\sin(x_\text{c}) - y_\text{c}}{\cos(\phi_n^\star) - x_\text{c}}
    \cdot (x - x_\text{c})
    + y_\text{c}
\right],
\quad \forall n \in N.
\end{equation}
The function representing the Lorentzian-constraint can then be defined as a function of phases \(\phi\) as 
\begin{equation}
\label{eq:27}
f(\phi) = \left(\frac{j + e^{j\phi}}{2}\right).
\end{equation}
Then for the given mapping center \((x_\text{c}, y_\text{c})\), the phases (\(\phi_n\)) of Lorentzian mapped points can be obtained  by solving the problem defined as
\begin{equation}
\label{eq:28}
\phi_{n} = \min_{\phi \in [0, 2\pi]} \left\| f(\phi) -l_n\left(\operatorname{Re}\left(f(\phi)\right)\right) \right\|
\quad \forall n \in N.
\end{equation}
\revised{Based on these equations, GMLCH can be expressed as an operator
\begin{equation}
\label{eq:gmlch}
\mathbf{q} = \mathcal{M}(\hat{\mathbf{q}}^\star; x_\text{c}, y_\text{c}),
\end{equation}
where the input is the set of ideal weights $\hat{\mathbf{q}}^\star$ and the output is the Lorentzian-constrained DMA weight vector $\mathbf{q}$. The operator $\mathcal{M}(\cdot)$ applies the mapping to each element $n \in N$ with respect to the chosen mapping center \((x_\text{c}, y_\text{c})\): first, the phase $\phi_n^\star$ of the ideal weight $\hat{q}_n^\star$ is used to solve the one-dimensional minimization problem in \eqref{eq:28}, yielding the Lorentzian-constrained phase $\phi_n$. The final DMA weights $\mathbf{q}$ are then constructed by substituting these phases into \eqref{eq:27}. Three specific cases of this generalized mapping, previously introduced, can now be examined.}

\subsubsection{Lorentzian-Constrained Phase Hologram}
\label{LCPH}
\revised{The first case corresponds to setting the mapping center at the origin, i.e., \(
\mathcal{M}(\hat{\mathbf{q}}^\star; x_\text{c}=0, y_\text{c}=0)\),
which yields LCPH as illustrated in Fig. \ref{fig:GMLCH_sub1}.} In this case, the phases of the Lorentzian-constrained DMA weights match those of the ideal weights, while the amplitudes vary depending on the phases and the elements in the lower half-plane are mapped to zero.
\subsubsection{Lorentzian-Constrained Euclidean Hologram}
\label{LCEH}
\revised{In LCEH, illustrated in Fig. \ref{fig:GMLCH_sub2}, GMLCH is applied with \(
\mathcal{M}(\hat{\mathbf{q}}^\star; x_\text{c}=0, y_\text{c}=0.5)\).} 
This mapping seeks for the optimal points on the Lorentzian-constrained circle that minimize the Euclidean distance to the ideal weights.
\subsubsection{Lorentzian-Constrained Unitary Shift Hologram}
\label{LCUSH}
\revised{With application of GMLCH via \(
\mathcal{M}(\hat{\mathbf{q}}^\star; x_\text{c}=0, y_\text{c}=1.0)\)}, the resulting mapping LCUSH, depicted in Fig. \ref{fig:GMLCH_sub3}, determines Lorentzian-constrained points (\(\mathbf{q}_n^\star\)) using the phases of the ideal weights (\(\mathbf{\hat{q}}_n\))  as \(\frac{j + e^{j\phi_n^\star}}{2}\). This corresponds to applying a unitary shift along the imaginary axis to the ideal weights. 

\subsection{Adaptive Radius Method for Lorentzian Mapping}

\begin{figure*}[!t]
\centering

\subfloat[]{\includegraphics[trim={0.0in 0.0in 0.0in 0.0in},clip,height=2.0in]{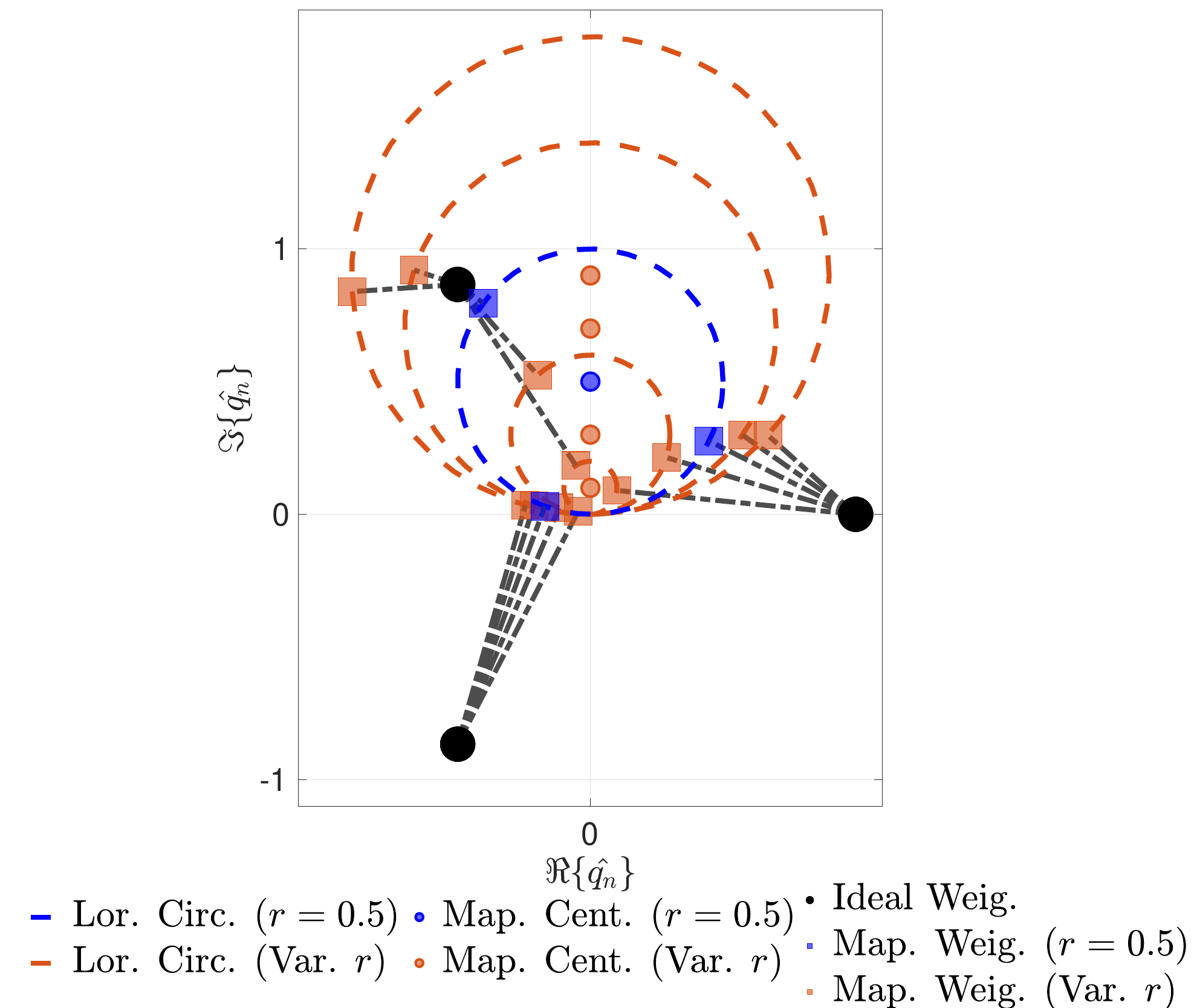}%
\label{fig:ARLCH_sub1}}
\hfil
\subfloat[]{\includegraphics[trim={0.0in 0.0in 0.0in 0.0in},clip,height=2.0in]{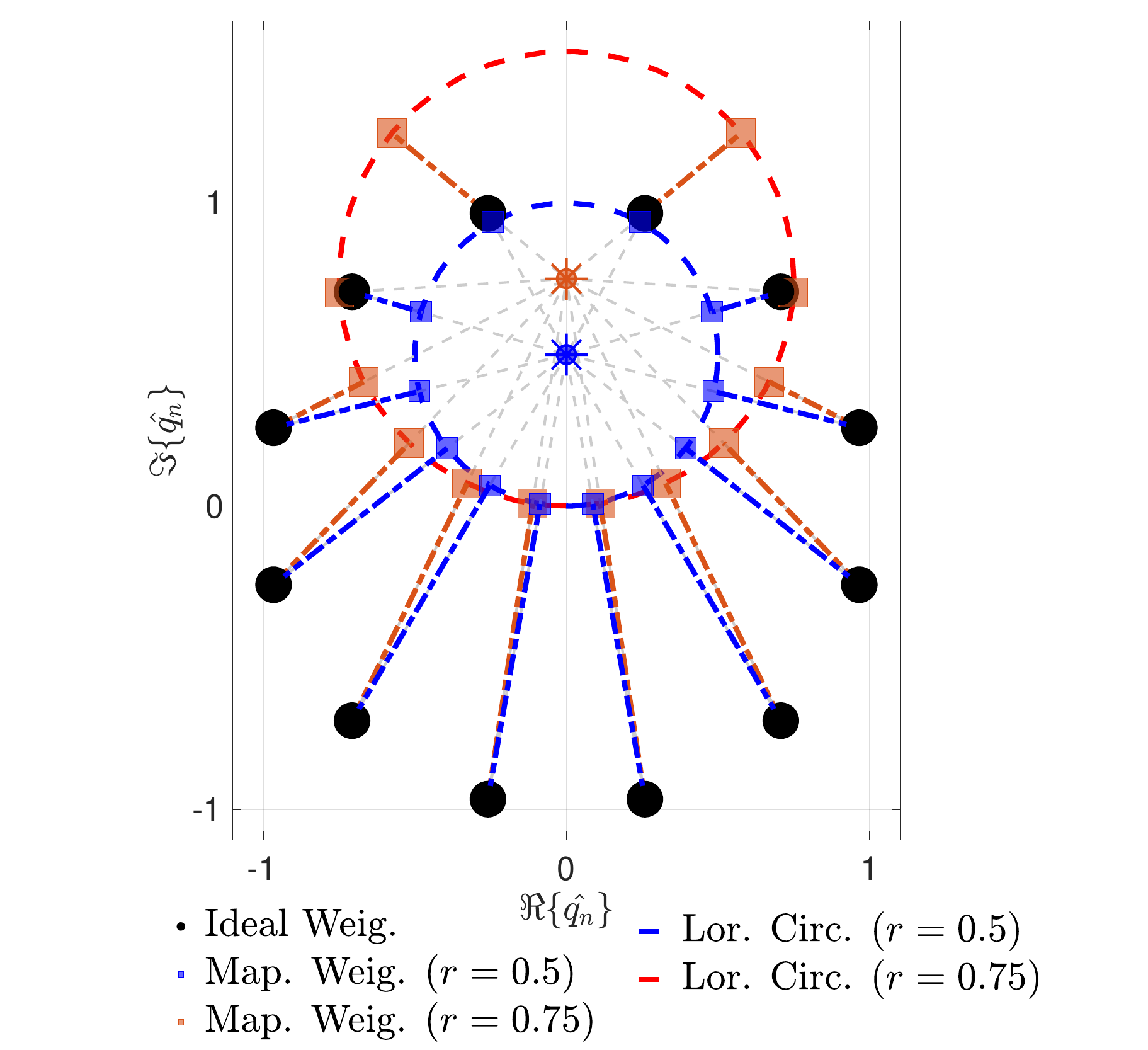}
\label{fig:ARLCH_sub2}}
\subfloat[]{\includegraphics[trim={0.0in 0.0in 0.0in 0.0in},clip,height=2.0in]{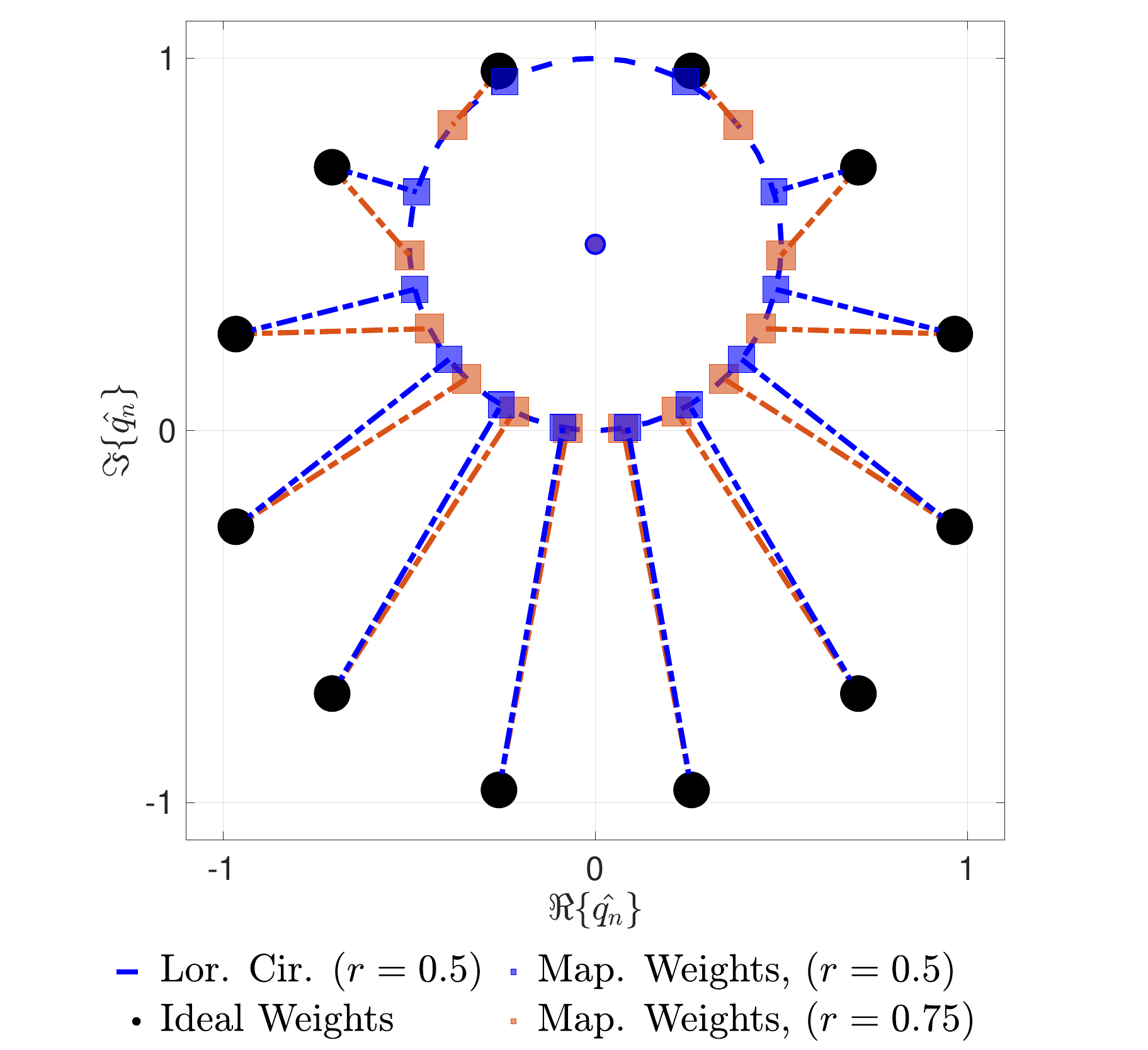}
\label{fig:ARLCH_sub3}}
\caption{(a) Lorentzian mapping with various diameters (\(D\)). (b) Comparison of Lorentzian mapping with \(D = 1.5\) against unitary Lorentzian mapping. (c) Final form of mapped weights in (b), where Lorentzian mapping with \(D = 1.5\) is normalized to satisfy the unitary condition.}

\label{fig:ARLCH_fig_ex}
\end{figure*}

A novel approach will be presented in this section, where the radius of Lorentzian-constrained circle (\(r\)) will be used as an additional DoF for achieving optimal mapping. Starting by revisiting \eqref{eq:27}, the equation for Lorentzian-constrained DMA weights can be defined in terms of the diameter of the Lorentzian circle, i.e., \(D=2r\), as
\begin{equation}
\label{eq:29}
f(\phi,D) = D\left(\frac{j + e^{j\phi}}{2}\right).
\end{equation}
In \eqref{eq:29}, \(D\) corresponds to the peak amplitude of tunable DMA weights and its value is determined by the intrinsic parameters of metasurfaces. As noted in the literature \cite{exp_dma1}, the extraction of \(D\) can be achieved through experimental setups or electromagnetic simulations based on the chosen metasurface element, though this is beyond the scope of this paper. In this paper, the tunability of the elements is constrained to a unitary Lorentzian circle, as defined in \eqref{eq:7p} with \(D = 1\). This constraint directs the optimization algorithm in substage 2 \eqref{eq:21} to find the DMA weights having amplitudes ranging from 0 to 1 and phases linked with them through the Lorentzian \revised{m}apping. On the other hand, restricting \(D\) to a specific value has also impact on the Lorentzian \revised{m}apping and relaxing it before mapping can lead to enhanced searching space for Lorentzian-constrained DMA weights (\(\mathbf{{q}} \in \mathbb{Q}^{N \times 1}\)) based on the ideal weights (\(\mathbf{{q}}^\star \in \mathbb{C}^{N \times 1}\)). Once the mapping is conducted with relaxed \(D\), the amplitudes of tunable weights can be normalized with it to ensure the unitary condition provided with \eqref{eq:7p}. But, relaxing \(D\) before Lorentzian mapping leads to different optimal points on Lorentzian circle, then performing optimization strictly on the Unitary Lorentzian-circle. This effect of value of \(D\) on the Lorentzian mapping is graphically illustrated in Fig. \ref{fig:ARLCH_fig_ex}, where the mapping is performed for the given \(D\) together with \((x_\text{c}, y_\text{c})=(0, D/2)\). 
Following the illustration of mapping of ideal weights onto Lorentzian circles with various diameters in Fig. \ref{fig:ARLCH_sub1}, the mapping for a specific diameter (\(D=1.5\)) is compared to the unitary Lorentzian circle mapping in Fig. \ref{fig:ARLCH_sub2}. The normalized form of the mapped weights for \(D=1.5\) is shown on the unitary Lorentzian circle in Fig. \ref{fig:ARLCH_sub3}. The comparison in Fig. \ref{fig:ARLCH_sub3} highlights that unitary and non-unitary mappings result in different optimal points on the unitary Lorentzian circle. 

Based on this idea, we present here a new mapping method, i.e., Adaptive Radius Lorentzian-Constrained Holography (ARLCH), where the optimization algorithm aim is to find optimal diameter for the Lorentzian-circle (\(D\)) along with the phases of tunable weights (\(\phi_n\)) so that the distance between DMA weights and ideal weights can be minimized. 
The mapping problem can be redefined using \eqref{eq:29} as,
\begin{equation}
\label{eq:30}
\min_{\boldsymbol{\Phi}, D} \left\|\mathbf{q}^\star - f(\boldsymbol{\Phi}, D) \right\|,
\end{equation}
where \(\boldsymbol{\Phi} = \{\phi_1, \phi_2, \dots, \phi_N\}\) is the set of angles for tunable \(N\) DMA elements. The optimization problem in \eqref{eq:30} differs from \eqref{eq:28}, which the minimization requires a multivariable optimization approach. To efficiently solve this problem with respect to the multiple variables \(\boldsymbol{\Phi}\) and \(D\), an AO strategy can be implemented, where each variable is optimized while keeping the other fixed. This approach ensures convergence to a local minimum and is commonly used in problems involving coupled variables.

When \(\boldsymbol{\Phi} = \{\phi_1, \phi_2, \dots, \phi_N\}\) is fixed in \eqref{eq:30}, the problem reduces to finding the projection of the unitary DMA weights, \(\hat{\mathbf{q}}\), onto the ideal weights \(\mathbf{q}^*\), where \(\hat{\mathbf{q}}\) is defined as:
\begin{equation}
\label{eq:31}
\hat{\mathbf{q}} = \frac{j + \exp(j \boldsymbol{\Phi})}{2}.
\end{equation}
This problem can be solved analytically, as described in the following lemma.
\newtheorem{lemma}{Lemma} 
\begin{lemma}
The optimal diameter of the Lorentzian circle, \(D^*\), which minimizes the distance between the ideal weights \(\mathbf{q}^*\) and the DMA weights under the unitary Lorentzian condition, \(\hat{\mathbf{q}}\), can obtained with:
\begin{equation}
\label{eq:32}
D^* = \frac{\operatorname{Re}(\hat{\mathbf{q}}^H \mathbf{q}^*)}{\hat{\mathbf{q}}^H \hat{\mathbf{q}}}.
\end{equation}
\textit{Proof:} See Appendix.
\end{lemma}

In the second stage of AO of \eqref{eq:30} for a given value of \(D = D^*\), the problem reduces to finding the optimal phase points \(\phi_n\) for a fixed diameter of the Lorentzian circle. This can be solved using a method similar to that in \eqref{eq:28}, by identifying the intersection points between the Lorentzian-constrained circle of diameter \(D_0 = D^*\), and the lines connecting the ideal weights \(q_n^\star\) to the center of circle at \(x_\text{c} = 0, y_\text{c} = D^*/2\). Given the ideal weights \(q_n^\star = x_n^\star + j y_n^\star\), where \(x_n^\star\) and \(y_n^\star\) are the real and imaginary parts, respectively, the equations for these lines are derived as:
\begin{equation}
\label{eq:33}
l_n(x) 
= x + j \left[ 
    \frac{y_n^\star - D^*/2}{x_n^\star} \cdot x 
    + \frac{D^*}{2} 
\right], 
\quad \forall n \in N.
\end{equation}

The optimal phase points on Lorentzian-circle of diameter \(D = D^*\) can then be determined by searching for the intersection points with the solution of following one-dimensional optimization problem:
\begin{equation}
\label{eq:34}
\phi_{n} =  \min_{\phi} \left\| f(\phi, d ) - l_n\left(\operatorname{Re}\left(f(\phi, d)\right)\right) \right\|,
\quad \forall n \in N,
\end{equation}
where \(\phi_n\) represents the optimal phase point for \(n\)-th element.
\begin{algorithm}[t]
\caption{Algorithm for ARLCH}
\label{alg:mapping_ARLCH}
\begin{algorithmic}[1]
\Require $\mathbf{q}^\star$ \eqref{eq:23} and its phases $\boldsymbol{\Phi}^* = \{\phi^*_1, \phi^*_2, \dots, \phi^*_N\}$, residual error $\epsilon$, maximum iterations $T_{\text{max}}$
\State Define cost function as $E(\boldsymbol{\Phi}, D) = \left\|\mathbf{q}^\star - f(\boldsymbol{\Phi}, D) \right\|^2$
\State Initialize $\boldsymbol{\Phi}^{(0)} \gets \boldsymbol{\Phi}^*$,  $E^{(0)} \gets 0$, 
\State Calculate $\hat{\mathbf{q}}^{(0)}$ using \eqref{eq:31} for $\boldsymbol{\Phi}^{(0)}$, 

\For{$t = 1, 2, \dots, T_{\text{max}}$}
\State Compute \( D^{(t)} \) for given \( \hat{\mathbf{q}}^{(t-1)} \), \( \mathbf{q}^\star \) based on \eqref{eq:32}.
\State Compute \( \boldsymbol{\Phi}^{(t)} \) for given \( D^{(t)} \)  by solving \eqref{eq:34},
\State Compute $\hat{\mathbf{q}}^{(t)}$ using \eqref{eq:31} for $\boldsymbol{\Phi}^{(t)}$, 
\State Compute \( E^{(t)} \gets E(\boldsymbol{\Phi}^{(t)}, D^{(t)}) \),
    \If{$|(E^{(t)} -E^{(t-1)} )/E^{(t)} | < \epsilon$}
        \State $D^f \gets D^{(t)}$, $\boldsymbol{\Phi}^f \gets \boldsymbol{\Phi}^{(t)}$, \( \hat{\mathbf{q}}^{(f)}  \gets \hat{\mathbf{q}}^{(t)} \)
        \State Update Algorithm \ref{alg:proposed_algorithm}, Step 7, \( \mathbf{\tilde{q}} \gets \hat{\mathbf{q}}^{(f)} \). 
        \State \textbf{Break.} 
    \EndIf
\EndFor
\end{algorithmic}
\end{algorithm}
Finally, solving the individual subproblems of \eqref{eq:30} in an alternating manner for fixed \( \boldsymbol{\Phi} \) and \( D \), using the solutions provided in \eqref{eq:32} and \eqref{eq:34}, leads to the computation of the Lorentzian-constrained DMA weights, denoted as \( \mathbf{\tilde{q}} \) in \eqref{eq:22}. The algorithm for this approach, which conducts the Lorentzian mapping with ARLCH, is presented in Algorithm \ref{alg:mapping_ARLCH}.  Note that the final form of DMA weights are obtained on the unitary Lorentzian circle with \eqref{eq:31} using the phases \(\boldsymbol{\Phi} \) optimized within the ARLCH (Step 11, \ref{alg:mapping_ARLCH}). Hence, \(\mathbf{\tilde{q}}\) adheres the relation defined with \eqref{eq:7p}.

\section{DISCUSSION AND COMPLEXITY ANALYSIS}
\subsection{Discussion over Proposed Algorithm}
\label{sec:discussion}
Having defined the techniques for optimizing DMA weights, the proposed beamforming algorithm in Algorithm \ref{alg:proposed_algorithm} can be summarized as an AO with three key stages: (i) determining the ideal (unconstrained) DMA weights $\mathbf{\tilde{q}}^\star \in \mathbb{C}^{N \times 1}$ (Steps 5--6), (ii) mapping these weights onto the Lorentzian-constrained set $\mathbf{\tilde{q}} \in \mathbb{Q}^{N \times 1}$ (Step 7), and (iii) computing the digital precoding vectors $\left\{\mathbf{w}\right\}_{m=1}^{M}$ (Steps 9--10). For GMLCH with a predefined mapping center, Step 7 of Algorithm~\ref{alg:proposed_algorithm} solves ~\eqref{eq:28}, whereas for ARLCH with a dynamic Lorentzian circle diameter, Algorithm~\ref{alg:mapping_ARLCH} is executed. As discussed previously, until now in the literature, the effect of mapping center on beamforming in wireless communication networks has not been investigated in detail for multi-user MISO/MIMO networks. The integration of GMLCH with SDP based approach presented in this paper for the AO of digital precoding vectors and DMA weights paves the way for flexibility of using different approaches such as LCPH , LCEH, LCUSH. We further propose a novel technique with ARLCH. This Lorentzian mapping approach enables an additional DoF with the relaxed diameter of Lorentzian-circle and it provides dynamic adjustments to the mapping center based on the distance between the ideal weights and the Lorentzian-constrained weight vectors, thereby enhancing beamforming performance compared to the GMLCH method. In the numerical results, we evaluate the performances of GMLCH and ARLCH in terms of power efficiency with different parameters such as SINR level, number of users, position of users. 

On the other hand, we implemented the method proposed in \cite{Yihan_Near} as a benchmark for optimizing DMA weights \((\mathbf{Q})\) with fixed digital precoding vectors. To optimize DMA weights using this technique, Steps 5–7 of Algorithm~\ref{alg:proposed_algorithm} are replaced by the approach presented in \cite[Algorithm 1, Step 11]{Yihan_Near}. 
In this technique, the optimization of Lorentzian-constrained DMA weights is carried out using ADMM, which solves two subproblems derived from the augmented Lagrangian. The first subproblem computes the ideal weights in the complex domain without any constant modulus constraints, aiming to minimize the transmitted power while satisfying the SINR requirements. The second subproblem projects these ideal weights onto the unit circle, minimizing a combination of the transmitted power and the deviation from the previously obtained ideal solution using augmented Lagrangian problem. Finally, the Lorentzian-constrained DMA weights are constructed by summing the imaginary j-term, associated with the Lorentzian condition, with the weights optimized on the unit circle. Based on that, one can expect this method to yield results similar to GMLCH \((x_\text{c}, y_\text{c})=(0, 1.0)\), i.e., LCUSH approach presented in \ref{LCUSH}. Hence, ADMM-SCA approach is utilized for benchmarking of LCUSH, serving as a baseline for our subsequent analysis on the impact of different mapping centers with GMLCH and the robustness of our proposed method (ARLCH) with flexible-diameter optimization.


\subsection{Complexity Analysis}

The complexity of proposed algorithm with Algorithm~\ref{alg:proposed_algorithm} is mainly driven by the complexity associated with solution of SDP problems using CVX toolbox. The worst-case complexity of the proposed algorithm can be determined in reference to the complexity of SDP using an interior point method, which is given with \( O(\max\{m, n\}^{4} n^{\frac{1}{2}} \log(1/\epsilon)) \), where \( m\) is the number of constraints, \( n \) denotes the number of optimization variables, and \( \epsilon \) represents the accuracy \cite{5447068}. Using this, the number of optimization variables can be determined with \( n = N_\text{r}\) for SDP problem in \eqref{eq:15} and \( n = N\) for SDP problem in \eqref{eq:21}, where the number of constraints (\( m\)) in both SDP problems  is equal to the number of users (\( K\)). Therefore, the overall complexity remains polynomial in both the problem size \( n \) and the number of constraints \( m\), ensuring computational feasibility.
For mapping DMA weights in GMLCH and ARLCH, the problem reduces to solving one-dimensional search problems, as defined in \eqref{eq:28} and \eqref{eq:34}, respectively. Each search has a complexity of \( \mathcal{O}(N) \), where \( N \) is the number of DMA elements.

Using the benchmarking method from \cite{Yihan_Near}, the worst-case complexity for optimizing DMA weights via the ADMM-SCA method is \( \mathcal{O}(N^{3.5} \log(1/\epsilon)) \). Although the SDP-based approach adopted in our algorithm requires a higher computational complexity, it offers greater flexibility, particularly in enabling optimization with a predefined mapping center and supporting various DMA weight optimization strategies. It is important to note that the primary objective of this paper is to investigate how the choice of mapping center and optimization technique affects overall performance within the MISO downlink network in terms of transmitted power efficiency. Designing more computationally efficient implementations of the proposed methods remains an important direction for future research.


\section{NUMERICAL RESULTS}

In this section, we provide numerical results for demonstrating the effectiveness of our proposed approach. First in Section~\ref{sec:single_user}, we investigate performance of the proposed LCH methods for single user scenario ($K=1$) and examine its robustness, and convergence in terms of global optimality against benchmarking methods and generic optimization problems. Besides LCH methods, we also solve the same scenarios by using FD-architecture, and DMA - Unrestricted weights. These problems can be used to investigate the optimality of SDP based approach for ($K=1$), in which both problems have the same DoF in terms of optimization variables and constraints. Then, in Section~\ref{sec:multiple_user}, we provide simulation results for multiple users scenario ($K>1$), compare LCH methods in different scenarios by also highlighting the effect of number of users on the performance gap between FD-based architectures and DMA-based architectures due to the DoF limitations associated with the reduced number of RF chains and Lorentzian-constraints.

For numerical simulations, we consider a downlink multi-user MISO cell, where BS equipped with DMA or FD-based architecture centered in the \textit{xy}-plane, and the users in the network are distributed in near-zone within a half-circular region defined by a radial distance range of \( 0.1d_\text{F} \leq \rho \leq 1 d_\text{F} \) and an angular span of \( -85^\circ \leq \theta \leq 85^\circ \).  Here, $d_\text{F}$ is the Fraunhofer distance, i.e., \(d_\text{F} \triangleq \frac{2L^2}{\lambda}\) for effective array length of \revised{\(L\)} and wavelength \({\lambda}\). \revised{The effective array size is given by \( L_\text{DMA} = \sqrt{[(N_e - 1)d_x]^2 + [(N_r - 1)d_y]^2} \), where \(N_e\) and \(N_r\) are the number of elements along the horizontal and vertical directions with inter-element spacings \(d_x\) and \(d_y\), respectively.}

Throughout the experimental study, we set the frequency as \( f = 28 \, \text{GHz} \), noise power as \( \sigma_k^2 = -75 \, \text{dBm} \). Optimization trials are performed to minimize the transmitted power of the BS while ensuring the minimum SINR requirement is satisfied for each user. In the numerical simulations, this requirement is defined uniformly across all users as \( \text{SINR}_{\text{min},k} = \delta \). For DMA, spacing of antenna elements \( d_x \) and RF chains \(d_y \) are set to \(\lambda/2\) for the initial studies, whereas the results for varied antenna spacing \( d_x \) are also provided. For FD, spacing of antenna elements in both directions  \( d_x \) and  \(d_y \) are set to \(\lambda/2\).

In the following simulations, we evaluate and compare five different LCH methods according to the method of optimization of DMA weights. The methods considered are as follows: 

\begin{enumerate}
    \item \textbf{LCPH:} Solved with SDR and GMLCH for \((x_\text{c}, y_\text{c}) = (0, 0)\),
    \item \textbf{LCEH:} Solved with SDR and GMLCH for \((x_\text{c}, y_\text{c}) = (0, 0.5)\),
    \item \textbf{LCUSH:} Solved with SDR and GMLCH for \((x_\text{c}, y_\text{c}) = (0, 1.0)\),
    \item \textbf{Baseline Method}: Solved with ADMM-SCA approach proposed in the reference paper \cite{Yihan_Near},
    \item \textbf{Proposed Method}: Solved with SDR and ARLCH.
\end{enumerate}

These methods are compared under identical conditions to ensure a fair evaluation of their effectiveness in optimizing DMA performance. It should be noted that, for all methods optimizations of digital precoding vectors are performed by solving problem \eqref{eq:15}.
\subsection{Single User}
\label{sec:single_user}
\begin{figure}[!t]
    \centering
    \subfloat[]{
        \includegraphics[width=0.40\textwidth]{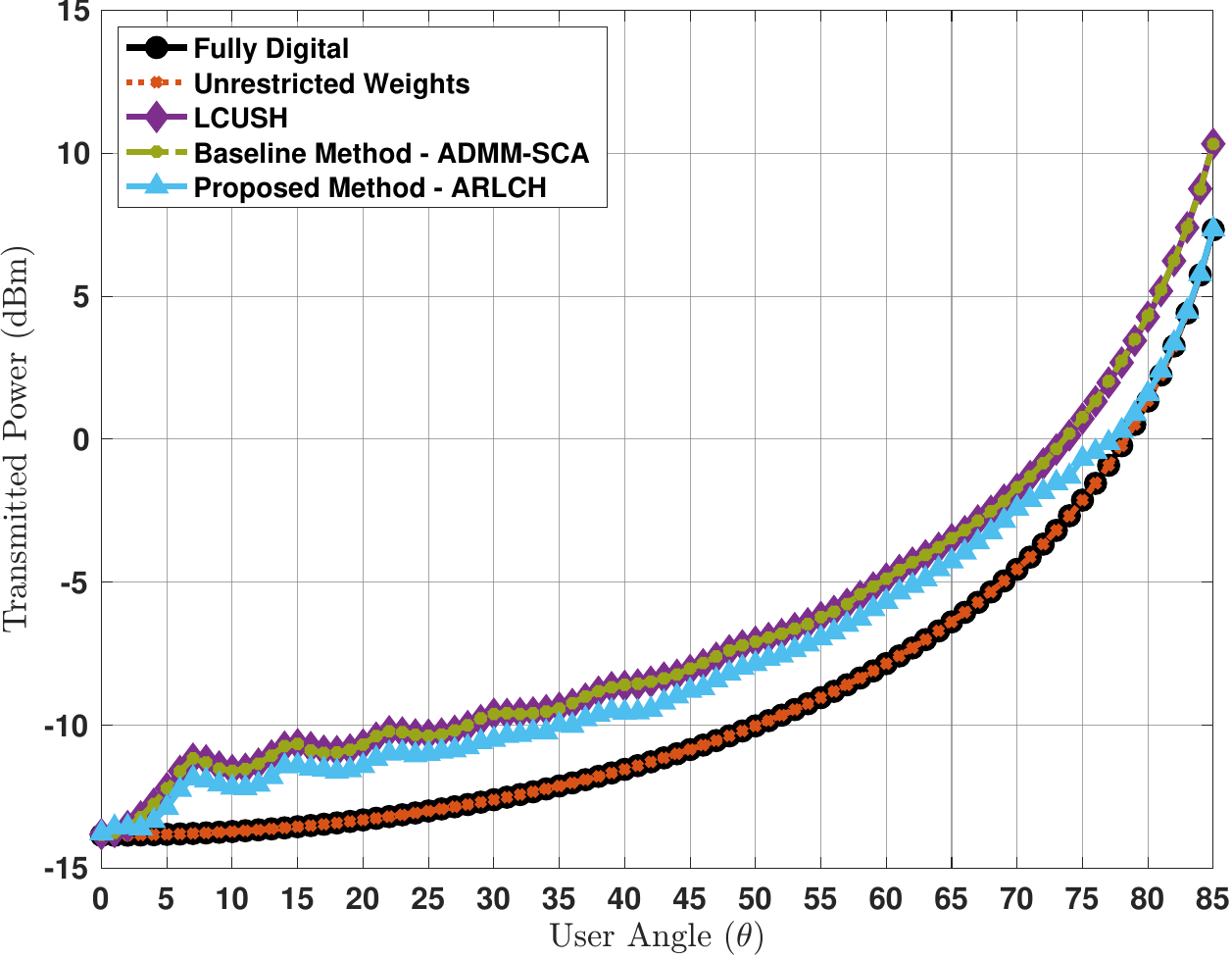}
        \label{fig:results_1a}
    }
    \hfill
    \centering
    \subfloat[]{
        \includegraphics[width=0.40\textwidth]{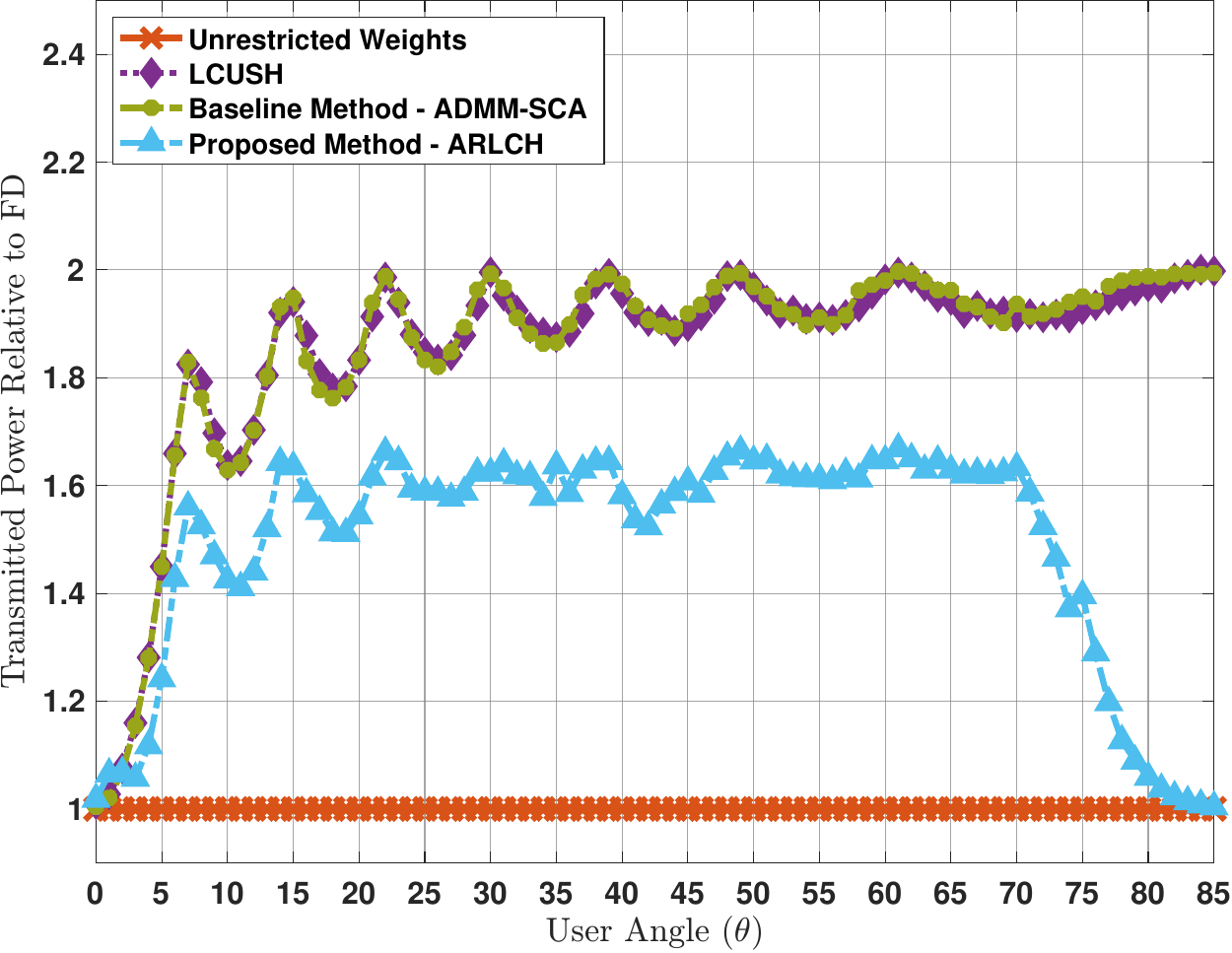}
        \label{fig:results_1b}
    }
    \caption{ Beamforming optimizations with different LCH techniques for \(\delta = 30\,\text{dB}\), \(K = 1\) and users located at \(\rho = 0.5d_\text{F}\) and \(\theta \in [0^\circ, 85^\circ]\). (a) Required transmit power versus user angle ($\theta$). (b) Transmit power relative to that of the FD case versus user angle ($\theta$).}
    \label{fig:results_1}
\end{figure}

In this section, we consider $K=1$ and assume identical frequency characteristics for all elements of DMA with \( \mathbf{H} = I_{N \times N} \). For both FD and DMA architectures, we consider a 16$\times$16 UPA with \(d_x=d_y=\lambda/2\). Here, number of RF chains and number of antenna elements per microstrip for DMA are set as \(N_\text{r}=16\) and \(N_\text{c}= 16\), respectively. \revised{When $K=1$, solving the DMA problem without Lorentzian constraints (DMA–Unrestricted weights) yields nearly the same nDoF as that of FD in terms of the optimization variables of the beamforming vectors. Hence, the results of the optimization problem for FD can be used to assess the optimality of the proposed approach under the idealized setting of the DMA setup with $\mathbf{H} = \mathbf{I}_{N \times N}$ \cite{aaltinoklu}.}

In the first study, we performed simulations with different LCH techniques, i.e., LCUSH, the baseline method ADMM-SCA, and the proposed method ARLCH for realizations of the single-user scenario with varied positions of the user in a fixed range of \(\rho = 0.5d_\text{F}\), while the angle of the users changes in \(\theta \in [0^\circ, 85^\circ]\) with $1^\circ$ resolution. Simulations are conducted for a minimum \revised{SNR} requirement of \(\delta_k = 30\,\text{dB}\).  In Fig.~\ref{fig:results_1a}, the minimized transmit power (\(P_{\text{Tx}}\)) achieved under different optimization setups is plotted as a function of the angle of the user. The results in Fig. ~\ref{fig:results_1} reveal that the transmitted power required to obtain 30 dB of \revised{SNR} for a single-user scenario in this setup increases when the position of the user gets broader with respect to the normal direction of DMA. This increase can be related to the smaller antenna gains of the individual elements for wider angles. Numerically, LCUSH performs very similar to the baseline method (ADMM-SCA) at all angles, where \(P_{\text{Tx}}\) increases from 0.04 \(\mathrm{mW}\)  to 10.8 \(\mathrm{mW}\)  as the angle changes from $0^\circ$ to $85^\circ$. However, within the same angular range, the proposed method (ARLCH) outperforms both methods, with \(P_{\text{Tx}}\) changing approximately from 0.04 \(\mathrm{mW}\) to 5.4 \(\mathrm{mW}\), and the performance of ARLCH is getting much better as the angle of the user increases. 

To further support the robustness of the proposed approach, the ratio of transmit power obtained using various DMA configurations to that of the FD case \(\left(P^{\text{DMA}}_{\text{Tx}} / P^{\text{FD}}_{\text{Tx}}\right)\) is  illustrated in Fig.~\ref{fig:results_1b} for the same simulation paramaters.  In these results, optimization trials for DMA-Unrestricted weights with transmit power values shown in Fig.~\ref{fig:results_1a} provide the same values (with  \(\left(P^{\text{DMA}}_{\text{Tx}} / P^{\text{FD}}_{\text{Tx}}=1\right)\)) for all angular positions, meaning that proposed algorithm with AO yields the global optimum solution when DoF limitations (due to Lorentzian constraints) are relaxed. On the other hand, when the available regions of DMA weights are reduced to the Lorentzian-circle with LCH methods, in all optimization methods of LCH, performance gap can be observed due to the phase and amplitude limitations introduced on the DMA weights. 
As depicted in Fig.~\ref{fig:results_1b}, the optimization of DMA weights through the adaptive adjustment of the Lorentzian-circle diameter in the proposed ARLCH method significantly reduces the performance gap compared to the other two approaches, i.e., LCUSH and the baseline method,  which both perform optimization over a fixed unitary Lorentzian circle. 
Numerically, the power ratio \(\left(P^{\text{DMA}}_{\text{Tx}} / P^{\text{FD}}_{\text{Tx}}\right)\) for ARLCH falls within the ranges \(1.00\text{--}1.42\) for \(\theta \in [0^\circ, 10^\circ]\), \(1.42\text{--}1.63\) for \(\theta \in [10^\circ, 70^\circ]\), and \(1.62\text{--}1.00\) for \(\theta \in [70^\circ, 90^\circ]\). In contrast, the power ratios for LCUSH and baseline method are nearly identical to each other and consistently higher than those for ARLCH rising from \(1.00\) to \(1.62\) in the region \(\theta \in [0^\circ, 10^\circ]\), and oscillating between \(1.62\) and \(2.00\) in the range \(\theta \in [10^\circ, 90^\circ]\). 

\revised{The performance gap trends in Fig.~\ref{fig:results_1b} stem from how DMA weight amplitudes are optimized. Although different optimization tools are used, both LCUSH and the baseline method (ADMM-SCA) first determine weights on constant-amplitude manifolds for phase alignment, the to satisfy the Lorentzian constraint these are adjusted by adding the $j$ term, without considering the amplitude profile of the Lorentzian circle. While the resulting phase mismatch is relatively small, both methods suffer from amplitude fluctuations caused by the Lorentzian circle’s sinusoidal profile, unlike the original constant-amplitude manifold. Because their projection strategies are similar, LCUSH and the baseline yield comparable results, highlighting both the robustness of our optimization tools and the impact of the mapping scheme on performance. In contrast, ARLCH adapts the Lorentzian diameter before projection, providing a balance between amplitude and phase matching, which enables direct control over the projected amplitudes and yields superior performance.}

\begin{figure}[!t]
    \centering
    \subfloat[]{
        \includegraphics[width=0.49\textwidth]{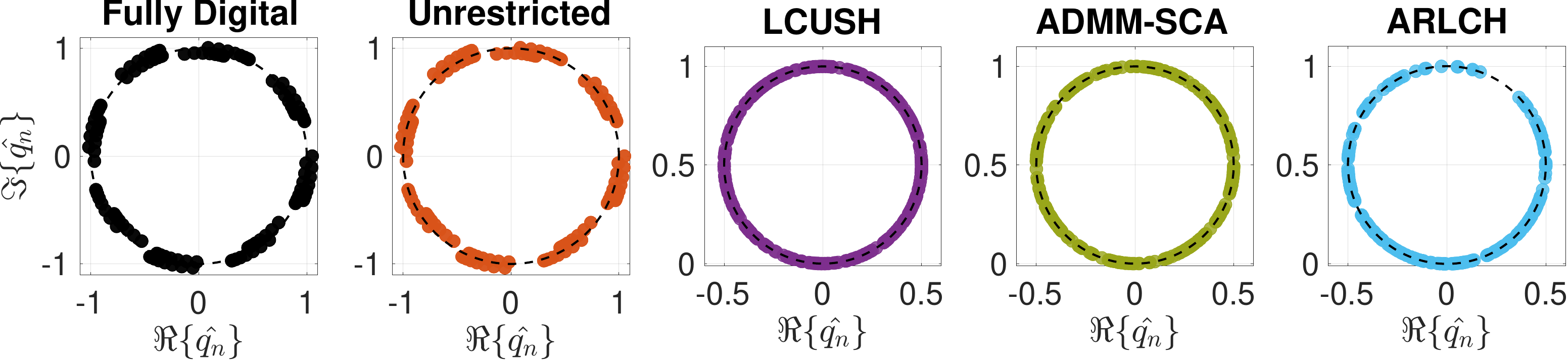}
        \label{fig:results_rev_1a}
    }
    \hfill
    \centering
    \subfloat[]{
        \includegraphics[width=0.49\textwidth]{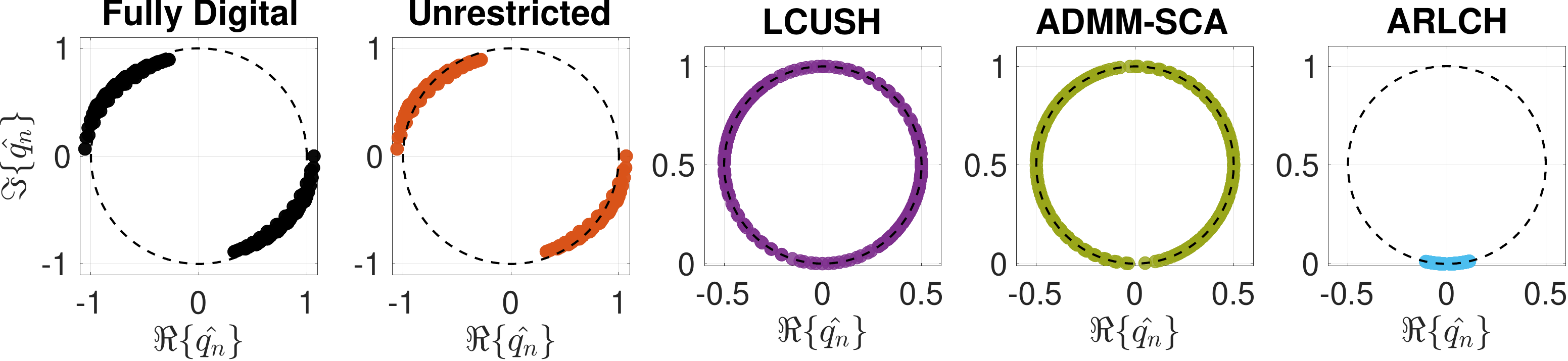}
        \label{fig:results_rev_1b}
    }
    \caption{Optimized beamforming weights with various LCH methods in single user scenario with \(\delta = 30\,\text{dB}\). (a) User position at $\rho=0.5d_\text{F}$, $\theta=50^\circ$. (b) User position at $\rho=0.5d_\text{F}$, $\theta=80^\circ$.}
    \label{fig:results_rev_1}
\end{figure}

\begin{figure}[!t]
    \centering
    \subfloat[]{
        \includegraphics[width=0.49\textwidth]{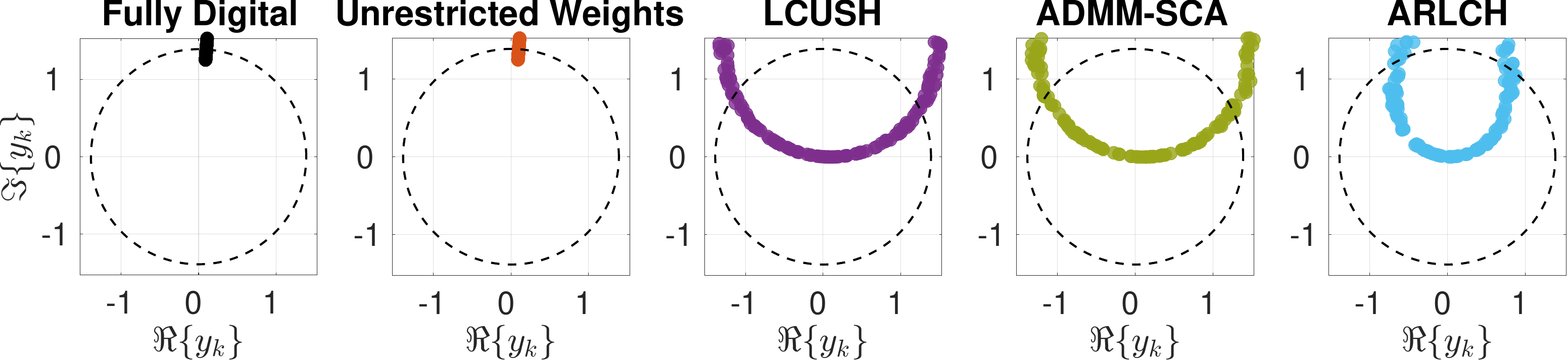}
        \label{fig:results_rev_2a}
    }
    \hfill
    \centering
    \subfloat[]{
        \includegraphics[width=0.49\textwidth]{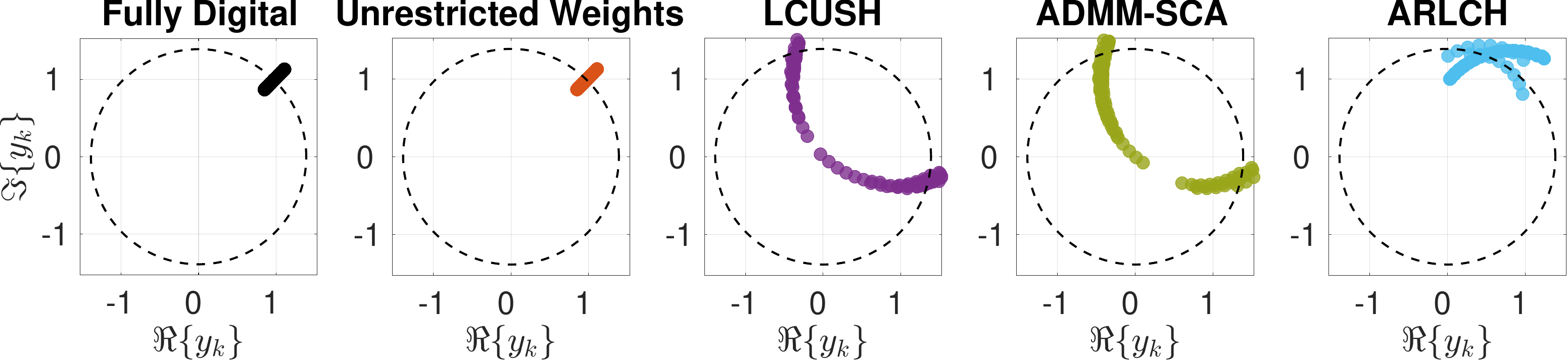}
        \label{fig:results_rev_2b}
    }
    \caption{Normalized received signal vectors at the user terminal with various LCH methods in single user scenario with \(\delta = 30\,\text{dB}\). (a) User position at $\rho=0.5d_\text{F}$, $\theta=50^\circ$. (b) User position at $\rho=0.5d_\text{F}$, $\theta=80^\circ$.}
    \label{fig:results_rev_2}
\end{figure}
\revised{To further support these observations, Fig.~\ref{fig:results_rev_1} and Fig.~\ref{fig:results_rev_2} show the distributions of the optimized DMA weights $\mathbf{Q}$ and the corresponding normalized received signal vectors, which refer to received signal contribution at the user terminal from each antenna element. For LCH schemes, the received signal vector is defined as $\mathbf{y} = \boldsymbol{\gamma}^H \circ (\mathbf{H}\mathbf{Q}\mathbf{w})$, and for FD as $\mathbf{y} = \boldsymbol{\gamma}^H \circ \mathbf{w}$, where $\circ$ represents element-wise product. Results are shown for UE1 ($\rho=0.5d_\text{F}$, $\theta=50^\circ$) and UE2 ($\rho=0.5d_\text{F}$, $\theta=80^\circ$), correspond to two key trends from Fig.~\ref{fig:results_1b}: ARLCH outperforms LCUSH and ADMM-SCA at both locations but matches FD and DMA-Unrestricted only at higher user angles.  
As shown in Fig.~\ref{fig:results_rev_1}, when beamforming requires a full-circle phase distribution, as in UE1, the projected weights on the Lorentzian circle for LCUSH and baseline method are spread across all $\phi_n \in [0, 2\pi]$, resulting in scattered element weights. From Fig.~\ref{fig:results_rev_2} we observe that ARLCH, by restricting amplitude fluctuations, confines the received signal vector elements closer to the optimal points of FD and DMA-Unrestricted weights, whereas larger magnitude variation of received signal elements in LCUSH and baseline method arise from magnitude variations in the complex DMA weights, which reduce constructive interference and power efficiency. While ideal beamforming relies mainly on phase alignment with minimal amplitude variation, the Lorentzian constraint in LCH-type schemes enforces a sinusoidal amplitude profile, inherently increasing these variations. ARLCH mitigates this effect by improving the alignment and power efficiency over the other two methods. For UE2, as shown in Fig.~\ref{fig:results_rev_1} for the FD and DMA-Unrestricted cases, the ideal beamforming weights are concentrated within specific angular regions, making them better-suited for amplitude control under the ARLCH scheme. By reducing the dynamic range of DMA weight amplitudes to $0$--$0.1$, ARLCH produces received signal vectors that are tightly clustered around the optimal phase points in Fig.~\ref{fig:results_rev_2}, yielding results almost identical to FD and DMA--Unrestricted cases due to minimized amplitude fluctuations. This shows the importance of amplitude control in achieving optimal beamforming performance. In contrast, because LCUSH and baseline method lack amplitude control, their optimized DMA weights span amplitudes from $0$ to $1$, resulting in larger-diameter received signal vectors, degraded constructive interference at the user terminal, and higher transmitted power compared to ARLCH.}



\begin{figure}[!t]
    \centering
    \includegraphics[width=0.4\textwidth]{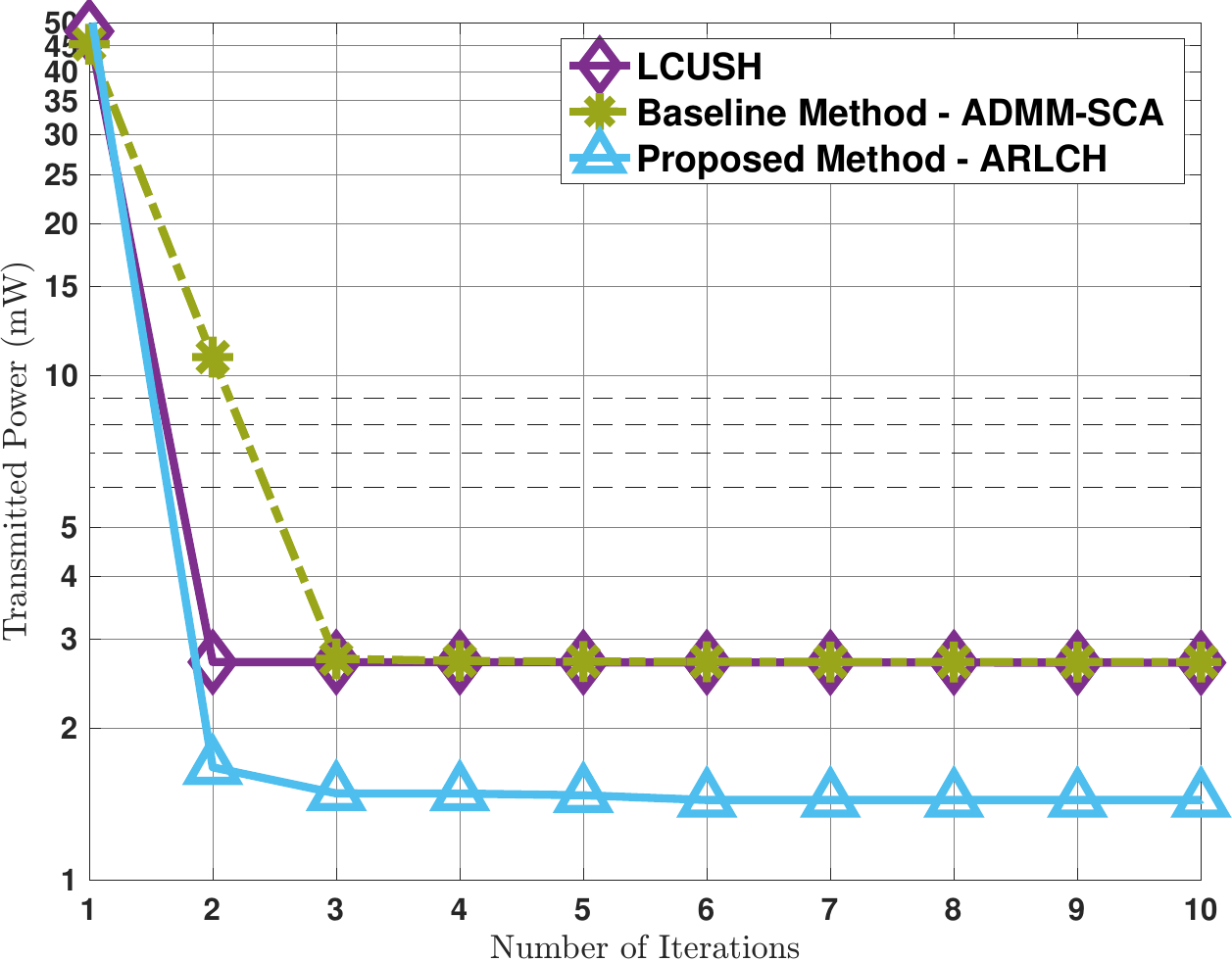}
    \caption{Convergence history of LCH methods for single user scenario with \(\delta = 30\,\text{dB}\) at  $\rho=0.5d_\text{F}$, $\theta=80^\circ$.}
    \label{fig:results_2}
\end{figure}

\begin{figure}[!t]
    \centering
    \begin{tikzpicture}
        \node[anchor=south west,inner sep=0] (image) at (0,0) {
            \includegraphics[width=0.4\textwidth]{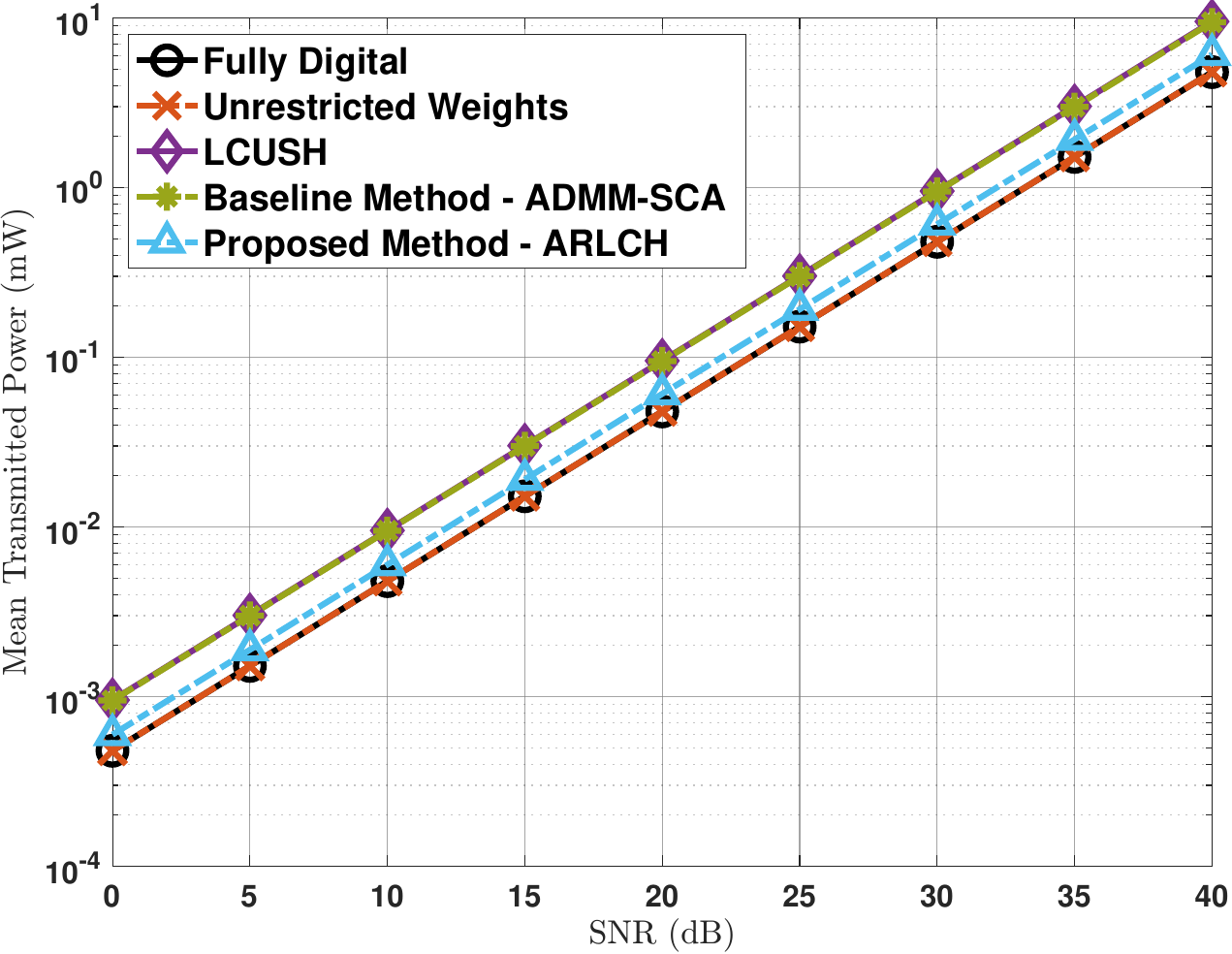}
        };

        \begin{scope}[x={(image.south east)},y={(image.north west)}]

            \begin{scope}[shift={(0.6, 0.20)}]
                \draw[thin] (0,0) rectangle (0.35,0.35);
                \node[anchor=south west,inner sep=0] at (0,0) {
                    \includegraphics[width=0.15\textwidth]{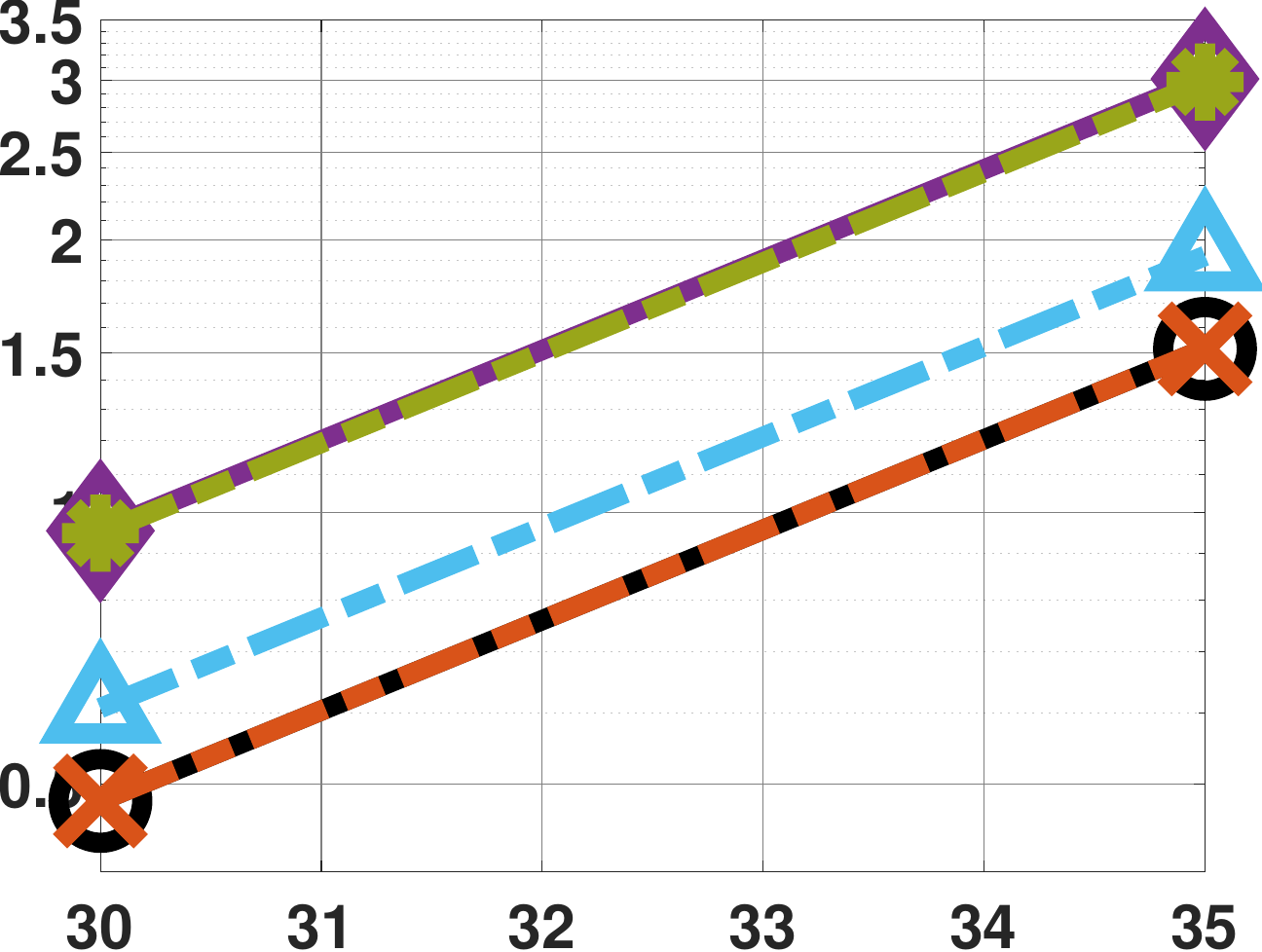}
                };
            \end{scope}
            \draw[gray,thin] (0.73,0.70) rectangle (0.9,0.92);

        \draw[gray, dotted, thick] (0.73, 0.70) -- (0.6, 0.55);
        \draw[gray, dotted, thick] (0.9, 0.70) -- (0.95, 0.55);
        \end{scope}
    \end{tikzpicture}
    \caption{\revised{Mean transmitted power versus minimum SNR requirement at $K=1$ for different LCH and benchmarking cases.}}
    \label{fig:results_3}

\end{figure}

Next, the convergence performance of the proposed method is illustrated in Fig.~\ref{fig:results_2}, in comparison with the other two LCH approaches for one of the simulations points of Fig.~\ref{fig:results_1}. All three LCH methods converge to their final values within a computationally reasonable number of iterations, achieving a residual error on the order of \(10^{-4}\) in fewer than 10 iterations. ARLCH outperforms LCUSH and baseline method in terms of the final transmitted power values \((P^{\text{DMA}}_{\text{Tx}})\), which highlights the effectiveness of the proposed approach. \revised{On the same hardware and tolerance settings, the wall-clock runtimes per instance were: LCUSH \(117.28\,\mathrm{s}\), baseline \(116.07\,\mathrm{s}\) and ARLCH \(119.25\,\mathrm{s}\); for ARLCH, approximately \(10\,\mathrm{s}\) of this total is spent in the inner AO loop for the Lorentzian-mapping step, which is reasonable given the observed transmit-power gains.}
Having shown the dependency of performance achieved with various LCH methods on the user positions, particularly angular sweep, the simulations are extended with 1000 realizations of Monte-carlo simulations for single user setup, where the randomly chosen positions of the user lie in the region defined with \( 0.1d_\text{F} \leq \rho \leq 1 d_\text{F} \) and \( -85^\circ \leq \theta \leq 85^\circ \). In Fig.~\ref{fig:results_3}, average of transmitted power values obtained at 1000 different positions of a single user as a function of required minimum \revised{SNR} values ranging from 0 to 40 dB are illustrated for ARLCH.
For all cases, the mean transmitted power increases linearly with increasing \revised{SNR}, where ARLCH always outperforms both LCUSH and baseline method across all \revised{SNR} values. Numerically, LCUSH and baseline method yield nearly identical mean transmitted power values, ranging from \(9.5 \times 10^{-4}\,\mathrm{mW}\) to \(9.5\,\mathrm{mW}\) as the \revised{SNR} increases from 0 dB to 40 dB. On the other side, the ARLCH method enables the performance gap relative to the FD case narrowing down, with mean transmitted power varying from \(6 \times 10^{-4}\,\mathrm{mW}\) to \(6.1\,\mathrm{mW}\) over the same \revised{SNR} range.


\subsection{Multiple Users}
\label{sec:multiple_user}

We demonstrated the robustness of proposed algorithm in \ref{sec:single_user} relative to the FD with the assumption of identical frequency characteristics of elements of DMA (\( \mathbf{H} = I_{N \times N} \)). In this section, we now consider the DMA elements with varying frequency selectivity, characterized by an attenuation constant $\alpha = 0.6\, \mathrm{m}^{-1}$ and a phase constant $\beta = 827.67\, \mathrm{m}^{-1}$, corresponding to the frequency response of a microstrip line fabricated on Duroid 5880 with a 30-mill substrate thickness at 28 GHz. Throughout this section, we present results for Monte-Carlo simulations with 1000 realizations for randomly distributed multiple users within the same region described in \ref{sec:single_user}.

\begin{figure}[!t]
    \centering
    \begin{tikzpicture}
        \node[anchor=south west,inner sep=0] (image) at (0,0) {
            \includegraphics[width=0.4\textwidth]{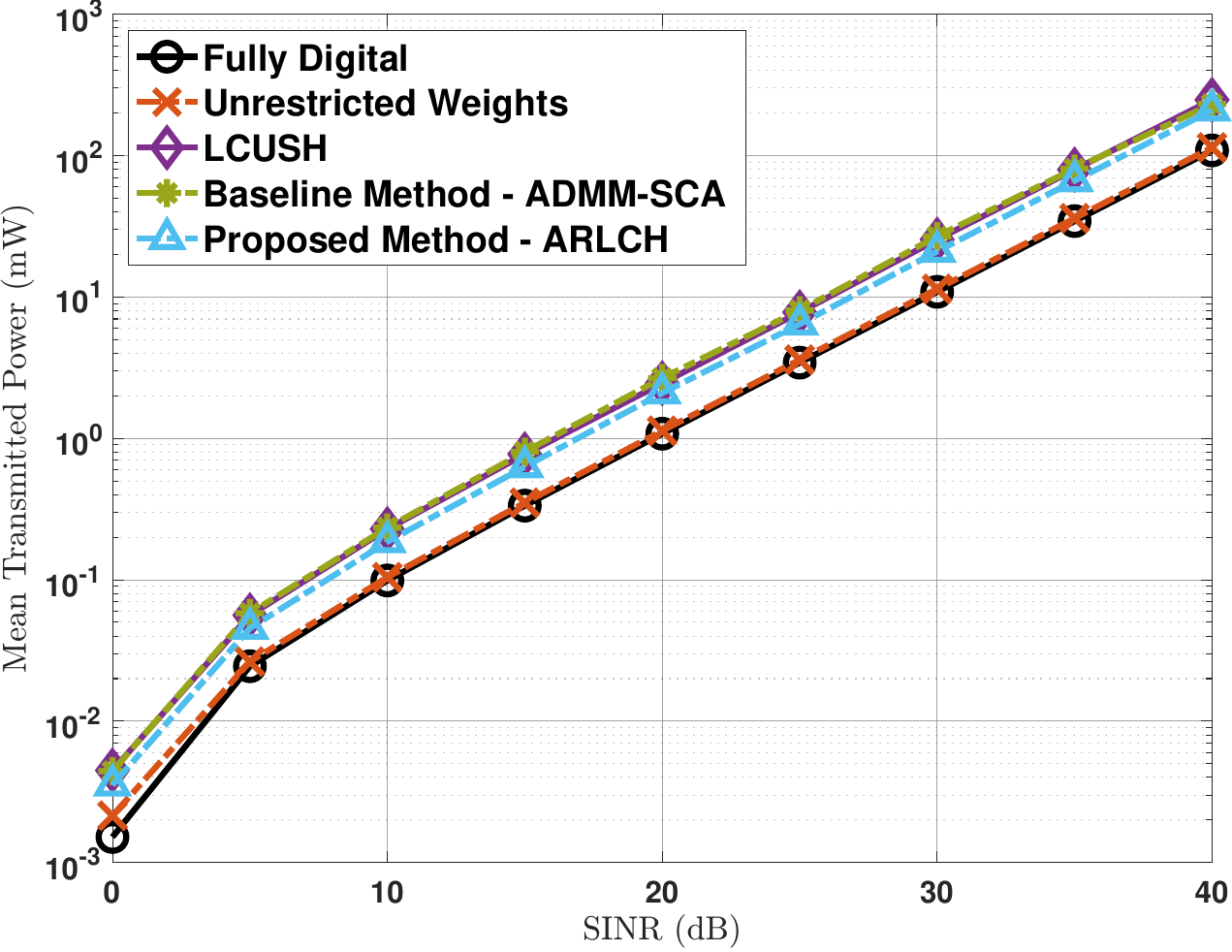}
        };

        \begin{scope}[x={(image.south east)},y={(image.north west)}]

            \begin{scope}[shift={(0.6, 0.20)}]
                \draw[thin] (0,0) rectangle (0.35,0.35);
                \node[anchor=south west,inner sep=0] at (0,0) {
                    \includegraphics[width=0.13\textwidth]{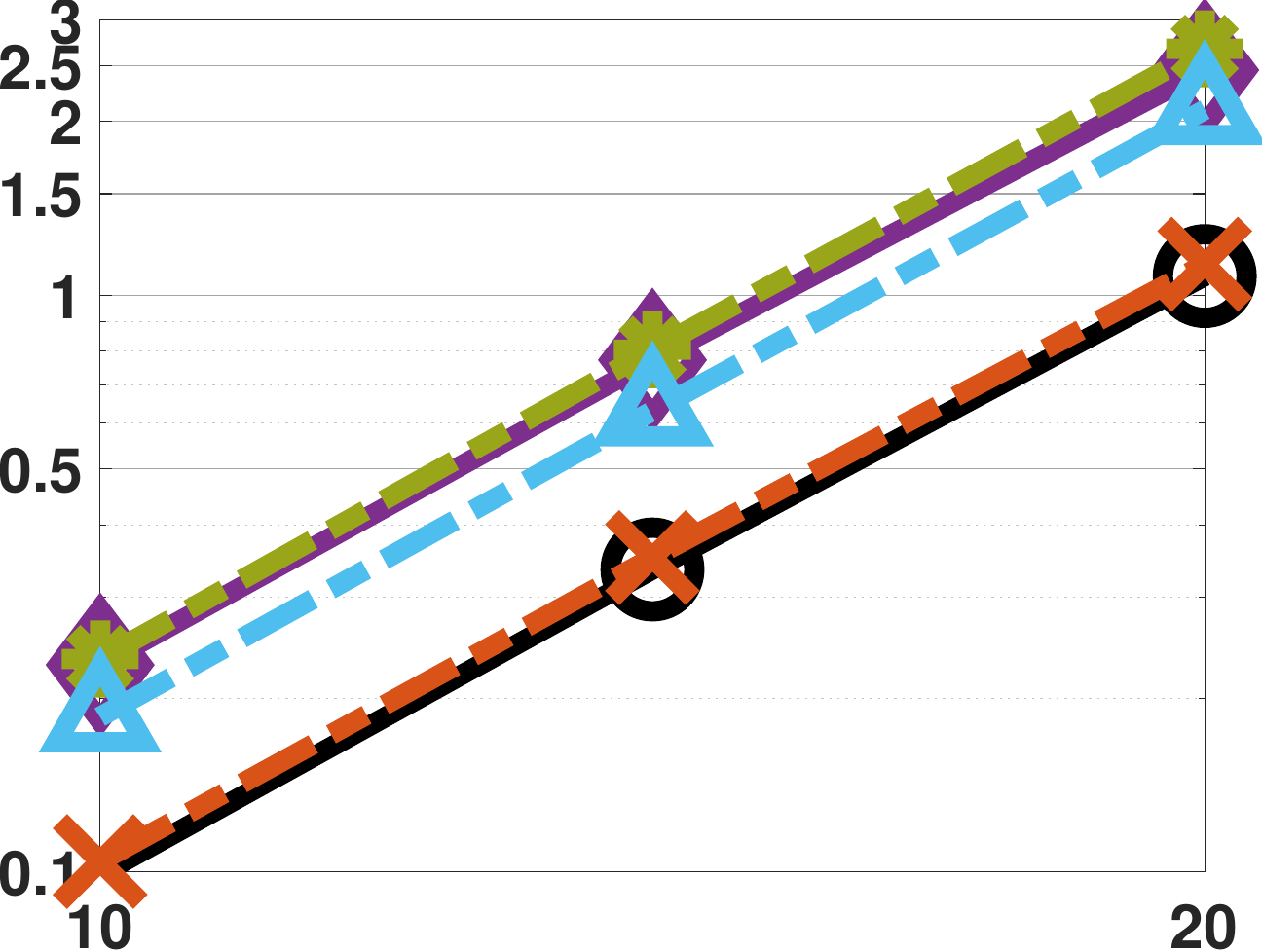}
                };
            \end{scope}
            \draw[gray,thin] (0.67,0.63) rectangle (0.86,0.88);

        \draw[gray, dotted, thick] (0.67, 0.63) -- (0.6, 0.55);
        \draw[gray, dotted, thick] (0.86, 0.63) -- (0.95, 0.55);
        \end{scope}
    \end{tikzpicture}
    \caption{Mean transmitted power versus minimum SINR requirement at $K=2$ for different LCH and benchmarking cases.}
    \label{fig:results_5}

\end{figure}

As for the initial studies, we consider a similar setup for both the DMA and FD architectures as used in the previous section, utilizing a \(16\times16\) UPA with inter-element spacing \( d_x=d_y=\lambda/2 \). In the first study, shown in Fig.~\ref{fig:results_5}, the average transmitted power over 1000 realizations of the two-user scenario (\(K = 2\)) is plotted as a function of the minimum SINR guarantees within the network. 
Similar to the \(K = 1\) scenario presented in the previous section, LCUSH and baseline method exhibit nearly identical performance, with average transmitted power increasing from approximately \(0.057\, \mathrm{mW}\) at \(\delta_k = 5\) dB to \(81\, \mathrm{mW}\) at \(\delta_k = 35\) dB. In contrast, the proposed ARLCH algorithm performs significantly better, with average transmitted power increasing from \(0.045\, \mathrm{mW}\) to \(66.0\, \mathrm{mW}\) as the SINR threshold increases from 5 to 35dB. The performance gap between ARLCH and the other two LCH-based approaches remains nearly constant across different SINR values, with ARLCH achieving approximately 20\% lower transmitted power in all cases with the same SINR requirements. 

\begin{figure}[!t]
    \centering
    \includegraphics[width=0.49\textwidth]{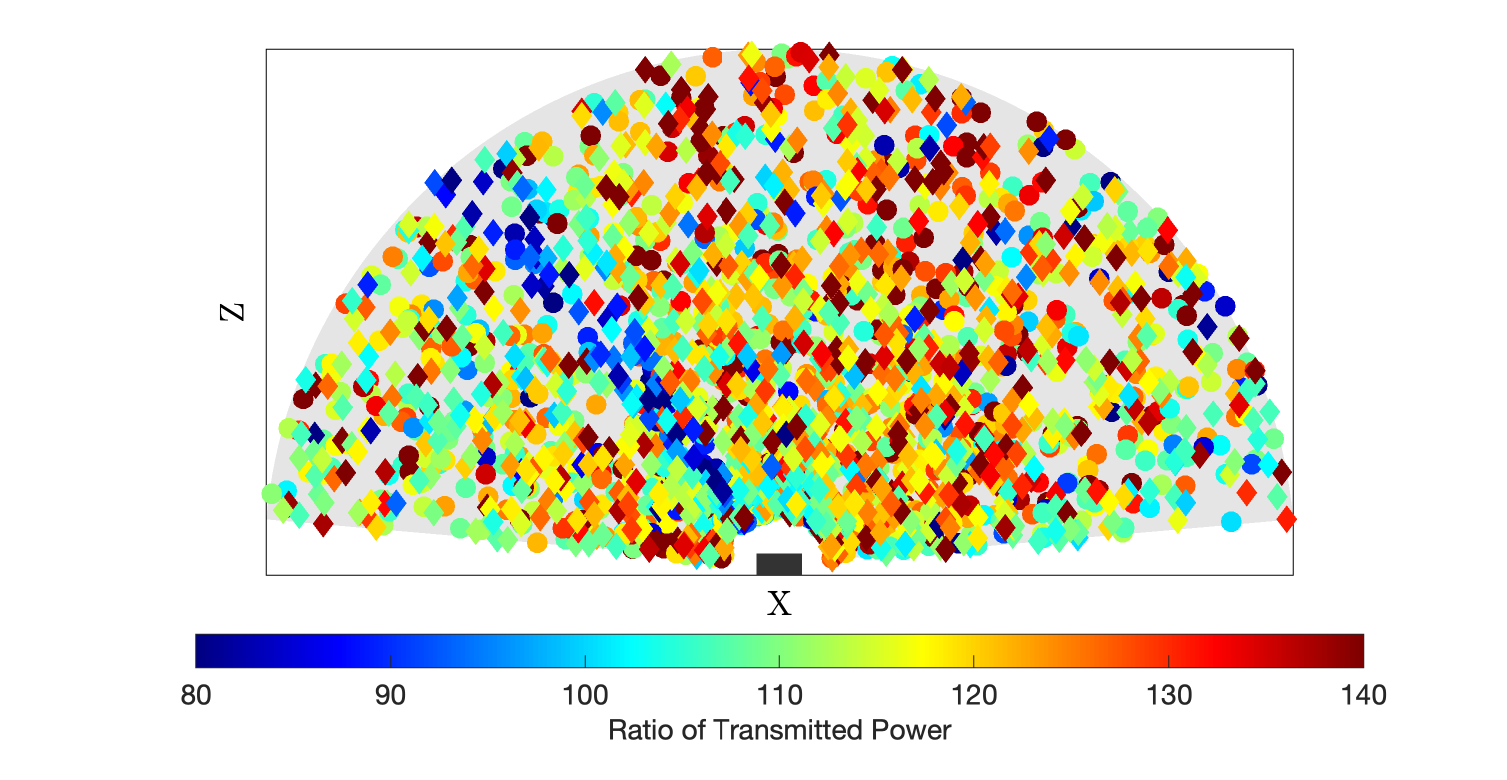}
    \caption{Ratio of transmitted power \(\left(P^{\text{LCUSH}}_{\text{Tx}} / P^{\text{ARLCH}}_{\text{Tx}}\right)\) over 1000 realizations with \(\delta_k = 30\, \text{dB}\) for all \(k\), and \(K = 2\). The realizations of User~1 are represented with circles, while those of User~2 are represented with diamonds.}
    \label{fig:results_6}
\end{figure}

To further investigate benefits of ARLCH, Fig.~\ref{fig:results_6} illustrates the ratio of transmit power required by LCUSH relative to the proposed method across 1000 independent realizations of randomly distributed users. The results reveal that ARLCH outperforms LCUSH in the majority of the realizations, with LCUSH requiring higher transmitted power to achieve the same SINR targets. These cases are illustrated with warm colors of colormap in the figure. In parallel with the observations for \(K = 1\) in Fig.~\ref{fig:results_1},  the performance difference between the two LCH-based methods is affected by the specific user locations.
Numerically, the transmit power ratio \(\left(P^{\text{LCUSH}}_{\text{Tx}} / P^{\text{ARLCH}}_{\text{Tx}}\right)\) is concentrated around 110\%--130\% for the vast majority of realizations, with some instances exhibiting even larger gaps, reaching up to 150\%.

\begin{figure}[!t]
    \centering
    \includegraphics[width=0.40\textwidth]{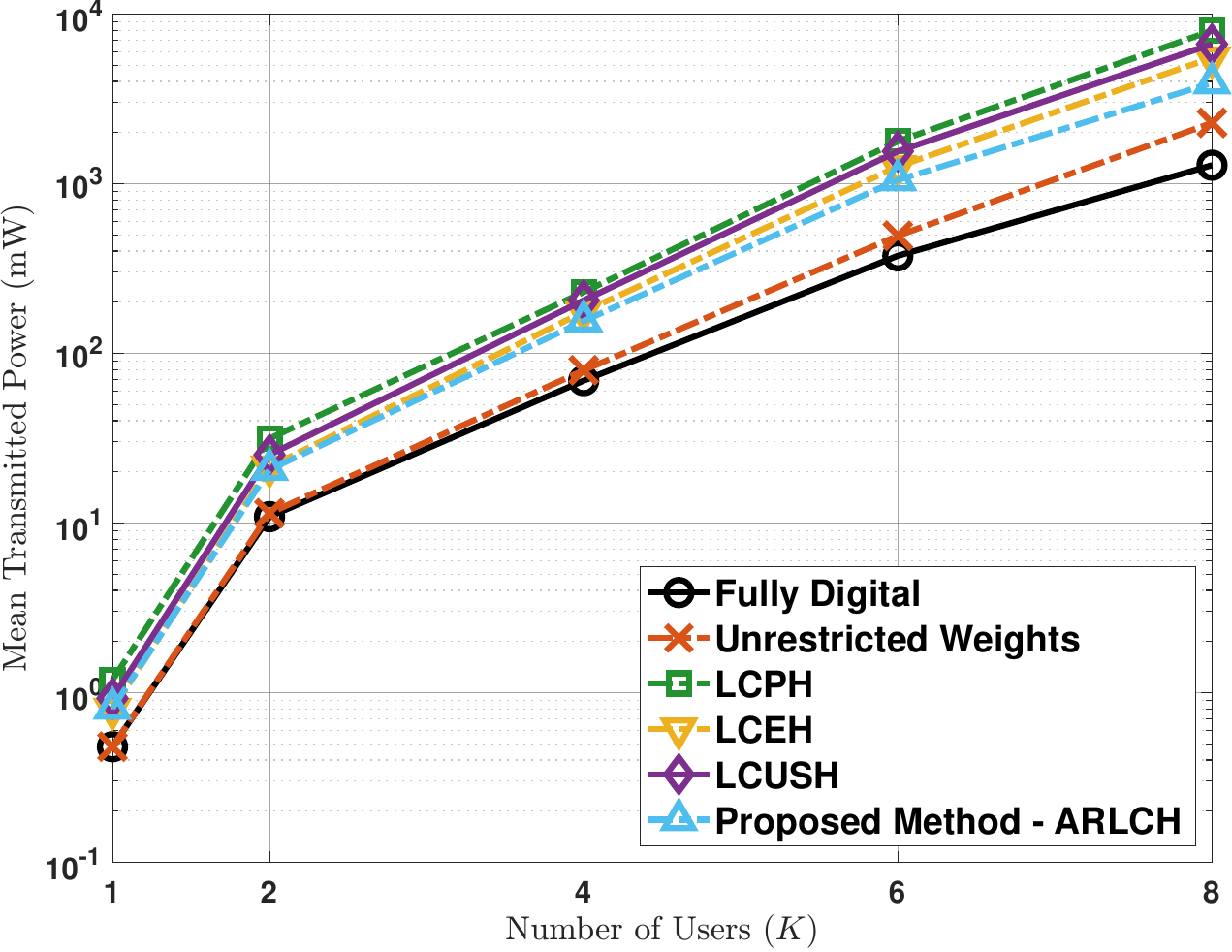}
    \caption{Mean transmitted power versus number of users with \(\delta_k = 30\, \text{dB}\).}
    \label{fig:results_7}
\end{figure}

Results presented up to now including both  \(K = 1\), and \(K = 2\) scenarios can also be examined for the comparison of LCUSH with the baseline method (ADMM-SCA). Our approach for achieving different types of LCHs in a generalized method, consisting of integration of SDR problems with GMLCH, achieves identical performance to the baseline ADMM-SCA method in terms of transmitted power, when the mapping center of GMLCH is set to \((x_\text{c}, y_\text{c}) = (0, 1.0)\) (LCUSH). These results validate our discussions in ~\ref{sec:discussion}. For the remainder of the paper, we proceed with the LCUSH method to achieve holographic beamforming for the mapping center \((x_\text{c}, y_\text{c}) = (0, 1.0)\), which, as detailed in Section~\ref{sec:Intro}, is one of the most commonly used LCH configurations in the literature.

Results in Fig. \ref{fig:results_3} and Fig. \ref{fig:results_5} can be also evaluated in terms of the performance gap between the LCH methods and the FD case. With \(K = 2\), this gap is observed to be larger than the one obtained in the single-user scenario. This is attributed to the reduced DoF of the DMA, as discussed in our previous study~\cite{aaltinoklu}. On the other hand, the proposed method in this paper enables reduction in this performance gap with an improvement within in the Lorentzian-mapping. To investigate this further, we performed Monte Carlo simulations with the same setup as in previous simulations as a function of number of users (\(K \)) within the network. For this analysis,  minimum SINR guarantee is selected as \(\delta_k = 30\, \text{dB}\) for all \(k \in K\). In parallel with the findings of~\cite{aaltinoklu}, the results presented in Fig.~\ref{fig:results_7} demonstrate that the performance gap between DMA and FD increases with the number of users. However, our results in here further reveal that this gap is also dependent on the specific LCH method utilized for achieving optimization of DMA weights. Notably, the proposed ARLCH scheme achieves superior performance as the number of users increases, outperforming all other techniques.  Moreover, ARLCH results in a smaller performance gap relative to the FD case than other methods, highlighting its scalability and robustness in multi-user scenarios.

Furthermore, the comparison of GMLCH techniques in Fig.~\ref{fig:results_7} provide valuable insight into the impact of the mapping center.  This factor has not been rigorously investigated in the existing literature. However, the results show that the performance of different LCH types differs significantly, which becomes even more prominent as the number of users increases. Numerically, LCUSH outperforms LCPH for all cases, achieving approximately 12.7\% and 16.8\% decrease in transmitted power for \(K > 6\) and \(K > 8\), respectively.
In all cases, LCEH always achieves lower transmitted power compared to both LCUSH and LCPH. The performance gap between LCEH and LCUSH increases with the number of users, indicating improved relative efficiency of LCEH in more demanding scenarios. Specifically, the transmitted power reduction achieved by LCEH relative to LCUSH scales from 13.9\% for \(K = 1\) to 17\% for \(K = 8\). As discussed in Section~\ref{sec:Intro}, LCUSH is the most commonly used LCH type in the literature; therefore, the superior performance of LCEH over LCUSH shown here is particularly noteworthy. 
On the other hand, the proposed ARLCH method achieves even better performance than LCEH, with the performance gain increasing as the number of users grows. Specifically, the transmit power reduction achieved by ARLCH relative to LCEH reaches 16.7\% for \(K = 6\) and 29.1\% for \(K = 8\). This comparison highlights the advantage of ARLCH in leveraging additional degrees of freedom introduced by the adaptive optimization of the Lorentzian-circle diameter prior to Lorentzian mapping, as proposed in this work.

\begin{figure}[!t]
    \centering
    \includegraphics[width=0.4\textwidth]{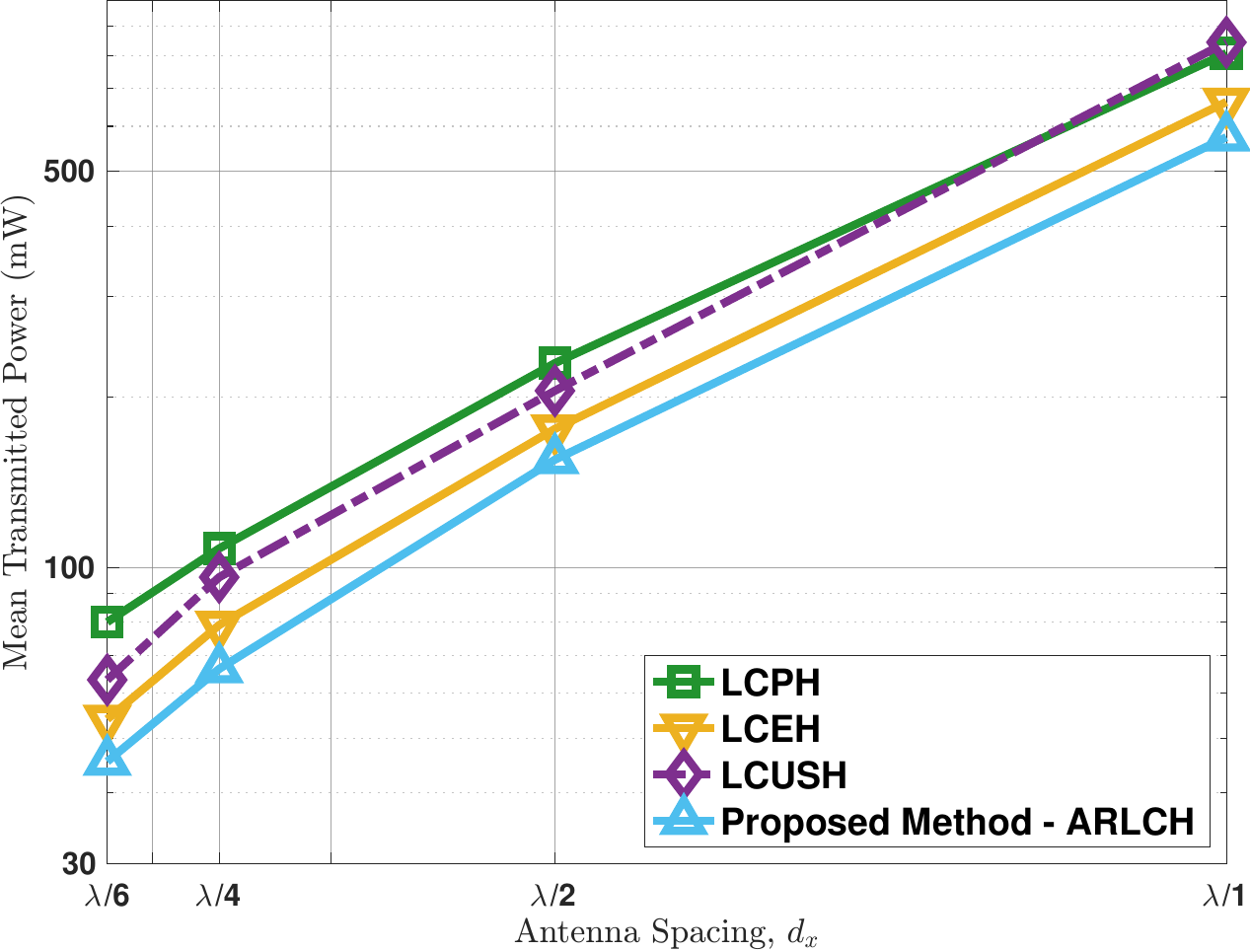}
    \caption{Mean transmitted power versus antenna spacing ($d_x$), and number of elements ($N_\text{e}$) at $K = 4$.}
    \label{fig:results_8}
\end{figure}

\revised{Until now, the effectiveness of ARLCH has been demonstrated for element spacing of $d_x = \lambda/2$. To further assess its performance, we now investigate denser element spacings, which are common in DMA designs. Thanks to high-dielectric microstrips and offset placement of metasurface elements, DMAs can physically support subwavelength spacings down to $d_x = \lambda/4-\lambda/6$ \cite{exp_dma1,em_dma2}. While mutual coupling is not explicitly modeled at the algorithmic level, it can be mitigated through hardware design, and it is often disregarded in DMA system-level optimization studies \cite{DMA_BF1,DMA_app4,Azarbahram_2024,Yihan_Near}. We therefore perform this analysis under an idealized setting where mutual coupling is not considered, consistent with prior works. 
Monte Carlo simulations were conducted for $K=4$ with a minimum SINR target of $30$~dB, considering four element spacings: $d_x = \lambda, \lambda/2, \lambda/4, \lambda/6$. With aperture size and RF chains ($N_\text{r}=16$) fixed, these correspond to array sizes of $16 \times 8$, $16 \times 16$, $16 \times 32$, and $16 \times 48$, respectively. Although denser antenna spacing increases the number of elements and DoF within a fixed aperture, it can also introduce grating lobes that must be suppressed. The results in Fig.~\ref{fig:results_8} show that transmit power decreases with denser spacing due to the higher DoF across all schemes, demonstrating the robustness of the proposed beamforming algorithm in handling these effects. However, the relative performance of different LCH schemes diverges under this setting. LCPH performs worst because, by design, it forces the phases of the DMA weights to match those of the ideal weights while clipping all elements in the lower half-plane to zero. This results in a higher fraction of non-radiating elements, and the degradation becomes more severe at denser spacings. LCUSH activates more elements and therefore outperforms LCPH, while LCEH achieves the lowest transmit power among GMLCH schemes due to its smoother magnitude profile. Importantly, ARLCH consistently outperforms all three GMLCH variants, with gains increasing from about $15\%$ at $d_x = \lambda$ to nearly $20\%$ at $d_x = \lambda/6$. This advantage stems from its adaptive mapping that better balances amplitude--phase relations, leading to lower grating lobes and reduced transmit power while maintaining the SINR targets. These findings confirm the robustness of ARLCH under denser DMA configurations and highlight the critical role of mapping strategies in practical system design.}

\section{Conclusion}
\revised{In this paper, we proposed a multiuser beamforming algorithm for DMA-assisted systems based on a unified Lorentzian-constrained holography (GMLCH) framework, which integrates LCPH, LCEH, and LCUSH within a single convex optimization setup for consistent benchmarking. Building on this, we introduced the Adaptive Radius Lorentzian-Constrained Holography (ARLCH), which dynamically adapts the Lorentzian circle radius and center to reduce projection mismatch. Numerical results demonstrate that ARLCH consistently achieves lower transmit power and superior scalability compared to other GMLCH schemes, narrowing the performance gap with fully digital architectures.
Beyond these contributions, several directions remain open for future work. A key step is the implementation on physical DMA hardware, where practical impairments, mutual coupling, and real-life limitations of tunability parameters of metasurfaces must be considered. Moreover, extending our framework to account for imperfect CSI and to support uplink transmission scenarios would strengthen its practical relevance. Another promising direction is the integration of our Lorentzian-constrained optimization with learning-based beamforming strategies, enabling real-time adaptation in complex propagation environments. Finally, given the increasing importance of sustainability, applying ARLCH to optimize energy efficiency in large-scale metasurface networks is a natural and impactful extension.}

\appendix[Proof of Lemma 1]
Starting with revisiting \eqref{eq:29}, based on \(\hat{\mathbf{q}}\) defined in \eqref{eq:31} 
\begin{equation}
\label{eq:ap_4}
f(\boldsymbol{\Phi}, D) = D\hat{\mathbf{q}},
\end{equation}
and using  \eqref{eq:ap_4} , the cost function in \eqref{eq:30} can be expanded as a function of \(D\) as:
\begin{align}
\label{eq:ap_1}
E(D) &= \| \mathbf{q}^* - D \hat{\mathbf{q}} \|^2 \nonumber \\
     &= (\mathbf{q}^* - D \hat{\mathbf{q}})^H (\mathbf{q}^* - D \hat{\mathbf{q}}) \nonumber \\
     &= (\mathbf{q}^*)^H \mathbf{q}^* - D (\mathbf{q}^*)^H \hat{\mathbf{q}} 
      - D \hat{\mathbf{q}}^H \mathbf{q}^* + D^2 \hat{\mathbf{q}}^H \hat{\mathbf{q}} \nonumber\\
     &= (\mathbf{q}^*)^H \mathbf{q}^* - 2D \operatorname{Re}(\hat{\mathbf{q}}^H \mathbf{q}^*) + D^2 \hat{\mathbf{q}}^H \hat{\mathbf{q}}.
\end{align}
Taking the derivative of \eqref{eq:ap_1} with respect to \( D\):
\begin{equation}
\label{eq:ap_2}
\frac{dE(D)}{dD} = -2\operatorname{Re}(\hat{\mathbf{q}}^H \mathbf{q}^*) + 2D \hat{\mathbf{q}}^H \hat{\mathbf{q}},
\end{equation}
and setting it to zero and optimal value for  \(D^*\) can be obtained as
\begin{equation}
\label{eq:ap_3}
D^* = \frac{\operatorname{Re}(\hat{\mathbf{q}}^H \mathbf{q}^*)}{\hat{\mathbf{q}}^H \hat{\mathbf{q}}}.
\end{equation}

\bibliographystyle{IEEEtran}
\bibliography{journal_abbreviations,references}

 




\vfill

\end{document}